%% file: jcap_article_magnification.tex
\documentclass[a4paper,11pt]{article}
\pdfoutput=1 

\usepackage{jcappub} 

\usepackage{graphicx}
\usepackage{booktabs}
\usepackage{amsmath,amsbsy}
\usepackage{amssymb}
\usepackage{amsthm}
\usepackage{subfig}
\usepackage{url}
\usepackage{listings}
\usepackage[normalem]{ulem}
\usepackage[]{natbib}

\newcommand{\be}{\begin{equation}}
\newcommand{\ee}{\end{equation}}
\newcommand{\ba}{\begin{eqnarray}}
\newcommand{\ea}{\end{eqnarray}}
\newcommand{\brr}{\begin{array}}
\newcommand{\err}{\end{array}}
\newcommand{\bc}{\begin{center}}
\newcommand{\ec}{\end{center}}

\newcommand{\lenstwo}{\lstinline!lenS!$^2$\lstinline!HAT!}
\newcommand{\stwo}{\lstinline!S!$^2$\lstinline!HAT!}
\newcommand{\lenspix}{\lstinline!LensPix!}
\newcommand{\healpix}{\lstinline!HEALPix!}
\newcommand{\camb}{\lstinline!CAMB!}
\newcommand{\nside}{\lstinline!NSIDE!}
\newcommand*\bell{\ensuremath{\boldsymbol\ell}}
\newcommand*\bellp{\ensuremath{\boldsymbol\ell^{\boldsymbol\prime}}}

\title{CMB weak-lensing beyond the Born approximation: a numerical approach}

\author[a,b,c]{Giulio Fabbian,}
\author[d,e,f]{Matteo Calabrese,}
\author[g,h,i]{Carmelita Carbone}
\affiliation[a]{Institut d'Astrophysique Spatiale, CNRS (UMR 8617), Univ. Paris-Sud, Universit\'{e} Paris-Saclay, B\^{a}t. 121, 91405 Orsay, France.}
\affiliation[b]{SISSA, Via Bonomea 256, I-34136, Trieste (TS), Italy.}
\affiliation[c]{INFN, Sezione di Trieste, Via Valerio 2, I-34127, Trieste, Italy.}
\affiliation[d]{Astronomical Observatory of the Autonomous Region of the Aosta Valley (OAVdA), Loc. Lignan 39, I-11020, Nus (AO), Italy.}
\affiliation[e]{INAF - Osservatorio Astrofisico di Torino, Strada Osservatorio 20, I-10025, Pino Torinese (TO), Italy.}
\affiliation[f]{INAF, Osservatorio Astronomico di Brera, via  E. Bianchi 46, I-23807, Merate (LC), Italy.}
\affiliation[g]{Universit\`a degli Studi di Milano, Dipartimento di Fisica, Via Giovanni Celoria 16, I-20133, Milano (MI), Italy.}
\affiliation[h]{INAF - Osservatorio Astronomico di Brera, Via Brera 28, I-20121 Milano (MI), Italy.}
\affiliation[i]{INFN - Sezione di Milano, via Celoria 16, I-20133 Milano (MI), Italy.}

\emailAdd{gfabbian@ias.u-psud.fr}
\emailAdd{matte.calabrese@gmail.com}

\abstract{
We perform a complete study of
the gravitational lensing effect beyond the Born approximation on the
Cosmic Microwave Background (CMB) anisotropies using a multiple-lens raytracing technique through cosmological N-body simulations of the DEMNUni suite. The impact of second order effects accounting for the non-linear evolution of large-scale structures is evaluated propagating for the first time the full CMB lensing jacobian together with the light rays trajectory. 
We carefully investigate the robustness of our approach against several numerical effects in the raytracing procedure and in the N-body simulation itself, and find no evidence of large contaminations. 
We discuss the impact of beyond-Born corrections on lensed CMB observables, and compare our results with recent analytical predictions that appeared in the literature, finding a good agreement, and extend these results to smaller angular scales. We measure the gravitationally-induced CMB polarization rotation that appears at second order, and compare this result with the latest analytical predictions. We then present the detection prospect of beyond-Born effects with the future CMB-S4 experiment. We show that corrections to the temperature power spectrum can be measured only if a good control of the extragalactic foregrounds is achieved. Conversely, the beyond-Born corrections on E and B-modes power spectra will be much more difficult to detect.  
}
 

\begin{document}
\maketitle
\flushbottom

\input{./intro.tex}

\input{./theory.tex}
\input{./algorithm.tex}
\input{./results.tex}

\input{./conclusions.tex}

\acknowledgments 
We thank Carlo Baccigalupi, Anthony Challinor, Enea Di Dio, Giuseppe Fanizza, Julien Grain, Stefan Hilbert, Antony Lewis and Giovanni Marozzi for useful discussion and comments. We thank Enea Di Dio, Giuseppe Fanizza , Giovanni Marozzi and Antony Lewis for providing the analytical results used for comparison with the simulations of this work. We thank Julien Carron for providing the fitting code for the Das \& Ostriker PDF model.\\* 
GF warmly thanks the hospitality of the University of Cambridge DAMTP and Institute of Astronomy where part of this work has been carried out. 
GF acknowledges the support of the Dennis Sciama Legacy fellowship and the CNES postdoctoral program. This work was partially supported by the RADIOFOREGROUNDS grant of the European Union Horizon
2020 research and innovation program (COMPET-05-2015, grant agreement number 687312).
MC and CC
acknowledge financial support to the ``INAF Fellowships Programme
2010'' and to the European Research Council through the Darklight Advanced
Research Grant (\# 291521). CC acknowledges the support from the grant MIUR PRIN 2015
``Cosmology and Fundamental Physics: illuminating the Dark Universe with Euclid''. MC has carried out part of the work for this project while supported by a grant (CUP B36G15002310006) of the European Union-European Social Fund, the Autonomous Region of the Aosta Valley and the Italian Ministry of Labour and Social Policy.\\*
This research used resources of the National Energy Research Scientific Computing Center (NERSC), a DOE Office of Science User Facility supported by the Office of Science of the U.S. Department of Energy under Contract No. DE-AC02-05CH11231.
Additional computing resources for this project have been provided by several class-C calls on the Fermi
and Marconi machines of the Centro Interuniversitario del Nord-Est per il Calcolo Elettronico (CINECA, Bologna,
Italy). The DEMNUni simulations were carried out on the Fermi machine at CINECA via the five million cpu-hrs budget provided by the Italian SuperComputing Resource
Allocation (ISCRA) to the class-A proposal entitled ``The Dark Energy and Massive-Neutrino
Universe". \\*
We acknowledge the use of the publicly available \lstinline!CAMB! code and \lstinline!HEALPix! package.

\appendix

\section{Weak lensing consistency relations}
\label{sec:appendix}
The full-sky magnification matrix can be written as
\be
\label{eq:magnstebbins}
\begin{split}
A_{ij} & = (1 - \kappa)\delta^K_{ij} - \gamma_{ij} + \omega \epsilon_{ij} \\
& = \delta^K_{ij} - \sum_{\ell m}\left( \psi_{\ell m} \nabla_i \nabla_j Y_{\ell m} +
\Omega_{\ell m} \epsilon^k_j \nabla_i \nabla_k Y_{\ell m} \right), 
\end{split}
\ee
where $\gamma_{ij}$ is a symmetric traceless tensor, 
\be
\gamma_{ij} = 
 \begin{pmatrix}
     -\gamma_1 & -\gamma_2  \\
  -\gamma_2 & \gamma_1
 \end{pmatrix} ,
 \ee
\noindent
$\omega \epsilon_{ij}$ is the anti-symmetric part of $A_{ij}$ and
$(1-\kappa)\delta^K_{ij}$ is the non-zero trace part. The last equality follows from the definition of Eq.~\eqref{eq:hs03}. 
Comparing the identities in Eq. \eqref{eq:magnstebbins} and
applying the relations between derivatives of scalar and spin-$s$
spherical harmonics of \cite{Hu2000}, we get 
\be
\kappa = \frac{1}{2} \sum_{\ell m }\ell (\ell +1)\psi_{\ell m}
Y_{\ell m}
\ee
\be
\omega = \frac{1}{2} \sum_{\ell m }\ell (\ell +1)\Omega_{\ell m}
Y_{\ell m}
\ee
\be
\boldsymbol\gamma\equiv \gamma_1 \pm i \gamma_2 = \frac{1}{2} \sum_{\ell m} \sqrt{\frac{(\ell
    +2)!}{(\ell -2)!}}\left( \psi_{\ell m} \pm i \Omega_{\ell m}\right)_{\pm 2}Y_{\ell m}.
\ee
\noindent
Being $\boldsymbol\gamma$ a spin-2 field, we can decompose it into E and B modes
and therefore the following relations between the harmonic coefficients hold:
\be
E^{\gamma}_{\ell m} \equiv \frac{1}{2} \sqrt{\frac{(\ell +2 )!}{(\ell -2)!}}\psi_{\ell m}, \qquad B^{\gamma}_{\ell m} \equiv \frac{1}{2} \sqrt{\frac{(\ell +2 )!}{(\ell -2)!}}\Omega_{\ell m}.
\ee
\noindent
We can then easily obtain the relations between convergence ($\kappa$),
rotation ($\omega$), shear E ($\epsilon$) and B-modes ($\beta$) and lensing and curl-potential angular power spectra \cite{Hu2000}:
\begin{gather}
\label{eq:angularspectra}
C^{\kappa\kappa}_{\ell} = \frac{1}{4} \ell^2 (\ell + 1)^2 C^{\psi\psi}_{\ell},\\
C^{\omega\omega}_{\ell} = \frac{1}{4} \ell^2 (\ell + 1)^2 C^{\Omega\Omega}_{\ell},\\
C^{\epsilon\epsilon}_{\ell} = \frac{1}{\ell^2 (\ell + 1)^2} \frac{(\ell+2)!}{(\ell-2)!} C^{\kappa\kappa}_{\ell},\\
C^{\beta\beta}_{\ell} = \frac{1}{\ell^2 (\ell + 1)^2} \frac{(\ell+2)!}{(\ell-2)!} C^{\omega\omega}_{\ell}.
\end{gather}
The cross-spectra $C^{\epsilon\beta}_{\ell}$, $C^{\omega\kappa}_{\ell}$, $C^{\kappa\beta}_{\ell}$, $C^{\omega\epsilon}_{\ell}$ and $C_{\ell}^{\psi^*\Omega}$ are all zero at first order in perturbation
assuming that the Universe is statistically parity invariant \cite{HirataSeljak2003}. Other cross spectra combination can be computed from the relations of harmonic coefficients provided above. Finally, note that at small scales the factor
\be
\label{eq:lfactor}
\lim_{\ell \to \infty}\frac{1}{\ell^2 (\ell + 1)^2} \frac{(\ell+2)!}{(\ell-2)!} \approx 1,
\ee
tends to one, thus the $\kappa$ and $\epsilon$ and $\beta$ and $\omega$ describe the same quantity \cite{Stebbins1996, HirataSeljak2003,KrauseHirata10}.

\section{Partial derivatives of an arbitrary
  spin-$s$ field} \label{sec:app_derivative_spin}
  
Starting from spin-$s$ spherical harmonics we can define their even and odd combination as harmonics
\be
_{s}F^{\pm}(\theta,\phi)=\frac{_{s}Y_{\ell m}(\theta, \phi) \pm(-1)^{s}\ _{-s}Y_{\ell m}(\theta, \phi)}{2} \label{eq:fpmdef}
\ee
If we consider the case of a spin-1 field and dropping the $(\theta, \phi)$ dependency for clarity we can write an explicit expression for $_{1}F^{\pm}$ applying the spin raising and lowering operators defined in \cite{Zaldarriaga96}
\ba
_{1}F^{+}(\theta, \phi)&=&\frac{-1}{\sqrt{\ell(\ell+1)}}(\partial_{\theta}Y_{\ell m}(\theta, \phi))=\frac{-1}{\sqrt{\ell(\ell+1)}}(\partial_{\theta}P_{\ell m}(\theta)_{\ell m}e^{im\phi})\\
_{1}F^{-}(\theta, \phi)&=&\frac{m}{\sqrt{\ell(\ell+1)}}\frac{Y_{\ell m}(\theta, \phi)}{\sin\theta}=\frac{m}{\sqrt{\ell(\ell+1)}}\frac{P_{\ell m}(\theta)_{\ell m}e^{im\phi}}{\sin\theta} 
\ea
where $P_{\ell m}(\theta)$ are the scalar associated Legendre polynomials. In the following we dropped the $s$ index to identify scalar quantities. We recall that the harmonics $_{s}F_{\ell m}^{\pm}$ can be recast in terms of spin-s associated Legendre polynomials $_{s}P_{\ell m}(\theta)$
\be
_{s}F_{\ell m}^{\pm}(\theta, \phi)=_{s}\!P_{\ell m}^{\pm}(\theta)e^{im\phi}
\ee
where $_{s}P_{\ell m}^{\pm}(\theta)$ are odd and even combinations of the polynomials in analogy with Eq. \eqref{eq:fpmdef}.

We recall that for a spin-s complex field $_{s}\eta(\theta,\phi)=R(\theta,\phi)+iI(\theta,\phi)$ we have
\ba
R(\theta,\phi) &=& (-1)^{H}\sum_{\ell m}( _{s}F^{+}_{\ell m}(\theta,\phi)\ _{s}E_{\ell m} + i _{s}F_{\ell m}^{-}(\theta,\phi)\ _{s}B_{\ell m})\\
I(\theta,\phi) &=& (-1)^{H}\sum_{\ell m}( _{s}F^{+}_{\ell m}(\theta,\phi)\ _{s}B_{\ell m} - i _{s}F_{\ell m}^{-}(\theta,\phi)\ _{s}E_{\ell m})
\ea
where H defines the arbitrary right-handed and left handed polarization convention. We assume H=1 in the following.
In case of spin-1 field and assuming harmonic coefficients of Eq. \ref{eq:deflection-field-calculation} we get
\ba
R(\theta,\phi)&=&-\sum_{\ell m}(_{1}P_{\ell m}^{+}(\theta)\psi_{\ell m}\sqrt{\ell(\ell+1)})e^{im\phi}=\sum_{\ell m}\partial_{\theta}\psi_{\ell m}Y_{\ell m}\equiv \partial_{\theta}\psi(\theta,\phi)\label{eq:dtheta}\\
I(\theta,\phi)&=&i\sum_{\ell m}(_{1}P_{\ell m}(\theta)^{-}\psi_{\ell m}\sqrt{\ell(\ell+1)})e^{im\phi}\\\nonumber
&=&\sum_{\ell m}\frac{im}{\sin\theta}P_{\ell m}(\theta)\frac{1}{\sqrt{\ell(\ell+1)}}\sqrt{\ell(\ell+1)}\psi_{\ell m}e^{im\phi} \equiv\frac{1}{\sin\theta}\partial_{\phi}\psi(\theta,\phi)\label{eq:dphi}
\ea
thus the derivatives of this spin-1 field give
\ba
\partial_{\theta}R(\theta,\phi)&=&\frac{\partial^{2}}{\partial{\theta}^{2}}\psi(\theta,\phi)\\
\partial_{\phi}R(\theta,\phi)&=&\frac{\partial^{2}}{\partial\phi\partial\theta}\psi(\theta,\phi) \label{eq:dphidtheta}\\
\partial_{\theta}I(\theta,\phi)&=&\frac{\partial}{\partial{\theta}}\left(\frac{1}{\sin\theta}\partial_{\phi}\psi(\theta,\phi)\right)\\
\partial_{\phi}I(\theta,\phi)&=&\frac{1}{\sin\theta}\frac{\partial^{2}}{\partial\phi^{2}}\psi(\theta,\phi) \label{eq:d2phi}
\ea
These are the expression of the partial derivatives of a spin-1 field computed by the \stwo\ routines used in this work. If the quantities in Eq. \ref{eq:dphidtheta} and \ref{eq:d2phi} are renormalized by an additional factor $1/\sin\theta$ we obtain the normalized version of the derivatives needed to compute the magnification matrix.

\section{Detection of polarization rotation in future experiments}\label{appendix:beta}
The direction dependent polarization rotation $\beta^{\rm rotation}$ mimics the effect of a direction-dependent cosmological birefringence process \cite{li2008, kamionkowski2009}, which generates coupling of CMB harmonic coefficients $E_{\ell m}$ and $B_{\ell^{\prime} m^{\prime}}$ with $\ell\neq\ell^{\prime}, m\neq m^{\prime}$. Thus, future experiments aiming at constraining parity violating processes - or primordial magnetic fields that can generate anisotropic polarization rotation in the CMB - might have to deal with the beyond-Born signal as a potential contaminant. The level of $\beta^{\rm rotation}$ expected from the raytracing does not seem to be of any concern. However, we checked if this was the case if $\beta^{\rm rotation}=\omega$. In order to compare the beyond-Born rotation signal with the expected sensitivity of future experiments to anisotropic birefringence processes, we converted the observed level of variance of our $\omega$ map into the power spectrum amplitude $K_{\alpha\alpha}$ of an equivalent scale invariant anisotropic birefringence process $\alpha$, i.e. $\ell(\ell+1)C_{\ell}^{\alpha\alpha}/2\pi = K_{\alpha\alpha} = {\rm constant}$. This class of models is commonly used in the literature to provide upper limits on anisotropic birefringence and can be easily related to the amplitude of a scale invariant primordial magnetic field that could source these kind of effect through Faraday rotation as \cite{de2013, pogosian2014}

\begin{equation}
B_{\textrm{1Mpc}} = 2.1 \times 10^{2}\left(\frac{\nu}{30\textrm{GHz}}\right)^{2}\sqrt{K_{\alpha\alpha}}\textrm{nG}.
\end{equation}
\noindent

In particular we found
\begin{equation}
K_{\alpha\alpha} = \frac{\sum_{\ell=30}^{5000}\frac{2\ell+1}{4\pi}C_{\ell}^{\omega\omega}}{\sum_{\ell=30}^{5000}\frac{2\ell+1}{2\ell(\ell+1)}} \approx 0.003\textrm{ deg}^{2} \qquad B_{\textrm{1Mpc}} \approx 4.8 \textrm{nG},
\end{equation}
\noindent
where we included in the calculation only the scales measured by our CMB-S4-like setup. The best constraint to date on $K_{\alpha\alpha}$ has been set by the POLARBEAR experiment \citep{pb-birefringence} who reported a null detection of $K_{\alpha\alpha}=0.33\textrm{ deg}^{2}$ and a 95\% confidence level upper limit of $K_{\alpha\alpha}\leq 1.0\textrm{ deg}^{2}$. CMB-S4 is expected to improve the error bars on $C_{\ell}^{\alpha\alpha}$ by almost three orders of magnitude \cite{cmbs4} and constrain $B_{\textrm{1Mpc}}\leq 0.6\textrm{nG}$ at 95\% CL. Thus, this level of signal might affect birefringence measurements in the upcoming years.\\* 
We also considered the possibility of accurately measuring $\beta^{\rm rotation}$ in this case through dedicated EB and TB anisotropic birefringence estimators \citep{kamionkowski2009, gluscevic2009} and found that the cumulative S/N of those will be around 2. However, recently proposed complementary techniques to measure rotation through radio galaxies polarization might also help in this process \cite{thomas2016}.

\bibliographystyle{JHEP}
\bibliography{multilens}

\end{document}

%% file: intro.tex

\section{Introduction}
The Cosmic Microwave Background (CMB) is now firmly established as one
of the most important cosmological probe, and has been spectacularly exploited to
very high sensitivity and accuracy by
the recent Planck satellite mission \cite{Planck2013_XVI,Planck2015_XIII} and current generation of suborbital experiments. In parallel to the studies on the primordial signal for cosmological applications,
the attention of the community has swiftly moved towards the
effects imprinted at later stages of the Universe
evolution on the CMB, the so-called secondary
anisotropies. The weak gravitational lensing of CMB anisotropies stands as the primary example of
such group and consists in a modification of the geodesic path of the
CMB light coming from the last scattering surface induced by growing matter
inhomogeneities.\\*
The first robust detections of such effect have been achieved using CMB temperature data only by ACT \cite{ACT} and SPT \cite{SPT, vanEngelen}, and later
confirmed by Planck with a significance greater than 25$\sigma$ \cite{PlanckXVII}. Only
recently, however, the evidence of lensing was detected for the first
time in polarization data by POLARBEAR \cite{pblens, pbbb}, SPTpol
\cite{hanson13}, ACTpol \cite{vanEngelen-actpol} and recently BICEP2/Keck Array \cite{keck_lensing}. Being sensitive to the whole matter distribution along the line of sight, CMB lensing can be used for cosmological analysis
to infer information about the Large Scale Structure
 (LSS) distribution and thus on the parameters that govern the physics of structure formation at intermediate and late time like (e.g. the dark
energy (DE) and massive neutrinos properties). Direct measurements of
CMB lensing can be improved and complemented by the cross
correlation analysis with
observations of the actual lenses in LSS surveys as 
independent tracers of the matter distribution. This approach has
already been exploited to obtained astrophysical and cosmological
information \cite{Bleem12, Sherwin12, act-chft, pb-cib,hanson13,Bianchini15, Giannantonio2015, Kirk15}. However, a major improvement is expected for this field in the forthcoming years when the next generation of high sensitivity CMB polarization experiments (Simons Array \cite{simons}, AdvACTpol \cite{advactpol}, SPT-3G \cite{spt3g} and ultimately CMB-S4 \cite{cmbs4}) together with the next generation of galaxy surveys DESI \cite{desi}, LSST \cite{lsst-sciencebook} and the ESA Euclid satellite \cite{Amendola13} will start observing the sky.\\*

The exponentially growing quality of CMB and galaxy surveys data requires a
great effort to simulate and make theoretical predictions for the cosmological
observables with the highest possible accuracy. In particular it will be crucial to take into account second-order effects and non-linear evolution of the large-scale structures at the same time. Several recent theoretical works \cite{Hagstotz14, pratten2016, marozzi2016, lewis-pratten2016, marozzi2016dec} studied the impact of the relaxation of the Born approximation to CMB lensing and lensed CMB anisotropies adopting the perturbative approach to the lens equation presented in
\cite{CoorayHu02, ShapiroCooray06, KrauseHirata10}. Similar studies have also been performed in the context of weak lensing (see e.g. \cite{petri2016, giocoli2016}). Analytical and numerical studies on the impact of the non-linear evolution of the matter distribution on several aspect of CMB lensing have also made significant progresses \cite{lewis-pratten2016, bohm2016, Namikawa2016, liu2016}.
At the numerical level, ray-tracing through large, high-resolution N-Body numerical
simulations is still the best tool we have to analyze the signal up to
the full non-linear scales. Despite the problem can be solved exactly \cite{Killedar11}, even though on limited sky fractions, a cheaper yet accurate and popular approach consists in extracting the 
lensing observables by photon ray-tracing along ``unperturbed'', i.e.
undeflected light paths in the so-called Born approximation
(e.g. \cite{Hilbert07a, Couchman99, Carbone09}). Approximated methods based on halo model formalism have also been recently proposed  in this context \cite{giocoli2017}.
In particular, \cite{Carbone13, carbone2016}
applied techniques based on the Born approximation to study a set of N-Body simulation with
different cosmologies and dark energy and/or massive neutrinos scenarios and investigate the variation of
the lensing pattern with respect to the standard $\Lambda$CDM model on the full sky. \\*
However, when facing the challenge of producing accurate and realistic simulations of the lensing effect, we must take
into account that each light ray undergoes several
deflections due to matter inhomogeneities. In this context, we should
therefore replace the modeling through a single effective deflection adopted in the Born approximation by
a multiple-lens (ML) approach. In the latter case, large volumes of matter are
projected onto a series of lens planes \cite{BlandfordNarayan, Jain00, Pace07, Hilbert09, Becker12, Fosalba08, petkova2014, petri-lenstool} so that the
continuous deflection experienced by a light ray is approximated by finite deflections at each of the mass planes. Faster approaches based on 
A ML algorithm for CMB lensing application were first sketched in \cite{DasBode}. In \cite{Calabrese14} we presented the first generalized implementation of this algorithm through large N-body simulations and applied it to study the lensing effect on both CMB temperature and polarization anisotropies.\\* 
Motivated by the recent theoretical results and to address concerns on the capability of our numerical setup to resolve beyond Born corrections in CMB lensing expressed in \cite{Hagstotz14}, 
in this work we expanded and improved the algorithm of \cite{Calabrese14}. In particular we modified the ray-tracing code to
explore and characterize the full set of lensing observables derived with the ML method up to
arcminute scales, where corrections and impact of non-linear effects are most noticeable and beyond the limit of \cite{Calabrese14}. \\
The paper is organized as follows. In Sec. \ref{sec:theory} we introduce
the theoretical background and notation used for our lensing
algorithm while in Sec. \ref{sec:algorithm} we discusse our ray-tracing technique emphasizing the improvements made from the previous version
of the algorithm. In Sec. \ref{sec:results} we discusse the properties of the lensing observables extracted from our simulation with particular emphasis on stability, accuracy and reliability of the
signal for different simulation setups. In Sec. \ref{sec:analytical-results-comparison} and \ref{sec:cmb-lensing} we compare the results of our simulations with analytical predictions for the lensing observables beyond Born approximation as well as their impact on lensed CMB observables and discuss perspective for the measurements of these corrections in future data sets. The last section draws the conclusions.

%% file: theory.tex

\section{Theory}\label{sec:theory}
\subsection{Gravitational light deflection}\label{sec:theory:grav}

In this Section we briefly introduce the relevant quantities of weak
lensing used in the rest of this paper. A detailed description of the weak
lensing formalism is summarized in several reviews
\cite{Bartelmann10,LewisChallinor06} and recent articles
\cite{Jain00,Hilbert09,DasBode,Becker12,Calabrese14} therefore we
refer the reader to these papers for further details.\\*
In weak lensing formalism, the effect of
deflections of light rays along the entire line of sight is described by the lens equation,
which maps the final position $(t, \boldsymbol\beta, \chi)$ of the
ray to the position of its source $\boldsymbol\theta$, i.e.
\be
\beta_i (\boldsymbol\theta, \chi) = \theta_i -
\frac{2}{c^2}\int_0^{\chi}\frac{f_K(\chi -
  \chi')}{f_K(\chi)f_K(\chi')}
\Psi,_{\beta_i}\left(\boldsymbol\beta(\boldsymbol\theta,
\chi'), \chi' \right) {\rm d}\chi', 
\label{eq:lenseq}
\ee
where $\Psi(t, \boldsymbol\beta, \chi)$ is a gravitational potential located on the photon path and $,_{\beta_i}$ its spatial derivative with respect to the photon's position $\boldsymbol\beta$.
Here $f_K(\chi)$ is the
standard angular diameter distance for a universe with curvature $K$.
The relative position of nearby light rays is quantified by the
derivative of the equation above
\be
\begin{split}
A_{ij}(\boldsymbol\theta,\chi) &\equiv \frac{\partial
  \beta_i(\boldsymbol\theta,\chi)}{\partial \theta_j}  = \\ 
& = \delta^K_{ij} - \frac{2}{c^2}\int_0^{\chi}\frac{f_K(\chi -
  \chi')}{f_K(\chi)f_K(\chi')}
\Psi,_{\beta_i\beta_k}\left(\boldsymbol\beta(\boldsymbol\theta,
\chi'), \chi' \right) A_{kj}(\boldsymbol\theta,\chi'){\rm d}\chi',
\end{split}
\label{eq:distortionmatrix}
\ee
where $\delta^K_{ij}$ is the Kronecker delta. 
The image distortions of light sources are described by the
magnification matrix\footnote{We will refer to the {\bf A} matrix as magnification matrix,  lensing jacobian or distortion tensor as synonyms in the following.} ${\bf A}(\boldsymbol\theta,\chi) \equiv
\left \{ A_{ij}(\boldsymbol\theta,\chi) \right \}$, which 
holds the information of the mapping induced
by lensing between the original image and the one at the current position on the lens plane. Note that the gravitational potential $\Psi$ is evaluated at the ray angular
position $\boldsymbol\beta(\boldsymbol\theta,\chi)$, while the
distortion itself - which is present at the r.h.s of
Eq.~\eqref{eq:distortionmatrix} and describes the {\it lens-lens coupling} - is computed at the
``background'' position $\boldsymbol\theta$.
The magnification matrix ${\bf A}$ is typically decomposed into four fields describing how the
light rays coming from a source at $\chi\equiv \chi_s$ are transformed by the passage through the matter distribution,
\be
A_{ij} \equiv \begin{pmatrix}
  \cos\omega & \sin\omega\\
  -\sin\omega& \cos\omega
\end{pmatrix} \begin{pmatrix}
  1 - \kappa -\gamma_1 & -\gamma_2 \\
  -\gamma_2 & 1-\kappa+\gamma_1
\end{pmatrix} \approx
\begin{pmatrix}
  1 - \kappa -\gamma_1 & -\gamma_2 + \omega \\
  -\gamma_2 - \omega & 1-\kappa+\gamma_1
\end{pmatrix},
\ee
where we assumed that the image rotation angle $\omega(\boldsymbol\theta,\chi_s)$ which defines the
rotation of the lensed image, is small, and we can work in the weak lensing regime. The
field $\kappa(\boldsymbol\theta,\chi_s)$ is referred to as the convergence while
$\gamma(\boldsymbol\theta,\chi_s) = \gamma_1(\boldsymbol\theta,\chi_s)
+ i\gamma_2(\boldsymbol\theta,\chi_s)$ defines the complex shear,
describing the shearing of the image along the two orthogonal directions of the basis, and can be decomposed into a
curl-free part, the shear E-modes $\gamma_{\epsilon}(\boldsymbol\theta,\chi_s)$ and a
divergence-free one, the shear B-modes $\gamma_{\beta}(\boldsymbol\theta,\chi_s)$. \\*
In the so-called Born approximation \cite{Carbone08,Hilbert09}, the
lens equation can be integrated over the unperturbed, unlensed
photon paths $\left(\boldsymbol\theta, \chi \right)$, therefore
dropping in Eqs.~\ref{eq:lenseq}, \ref{eq:distortionmatrix} the dependence over
the lensed position $\boldsymbol\beta$. While absent in first order weak lensing
approximation, shear B-modes are generated if we take into account the full
non-linear equation at higher order \cite{HirataSeljak2003,Hagstotz14}. Modern surveys \cite{Pen2002} can detect the presence of B-modes in the shear
field and although these are mainly
used as a monitor for systematic effects in weak-lensing data, this signal could also be used for cosmological applications such as e.g. testing anisotropic cosmological models \cite{uzan2016}.\\*
An additional assumption usually made is to consider negligible the coupling
between lenses at two different redshifts. Unlike the correction to the Born approximation, the lens-lens coupling
results in both B-mode generation and a net rotation of galaxy
images, which has also been measured in numerical
simulations \cite{Jain00}.\\*
The presence of these effects authorizes the introduction of an
auxiliary curl potential $\Omega(\boldsymbol\theta)^{\rm eff}$,
such that the deflection angle may be expressed as a combination of
the standard gradient contribution from a scalar field $\psi(\boldsymbol\theta)^{\rm eff}$, and a
curl contribution \cite{HirataSeljak2003}:
\be
\boldsymbol\beta (\boldsymbol\theta, \chi_{s}) = \boldsymbol\theta - \nabla \psi - \nabla \times \Omega, 
\label{eq:hs03}
\ee
where we defined the two-dimensional curl $(\nabla \times \Omega)_i =
\epsilon_{ij}\partial \Omega$ and dropped the spatial dependence of the potentials for sake of clarity. This potential $\Omega^{\rm eff}$ has to be
intended as an ``effective'' Born-like potential, integrated along the
line of sight, that encodes all the information about the rotation of the
image as curl-like patterns are originated in the signal through
multiple deflections. All the quantities above can be
treated on the sphere using the spin-$s$
spherical harmonic decomposition of the full-sky \cite{Hu2000,Becker12}. In
addition, combining results from \cite{Hu2000} and
\cite{Stebbins1996}, we can derive consistency relations between the components of the magnification matrix and the effective potentials (see Appendix~\ref{sec:appendix} for further details). 

\subsection{Multiple-lens-plane approach}
\label{sec:theory:multiple}
Following \cite{DasBode,Calabrese14}  we note that the Eqs.~\eqref{eq:lenseq},\eqref{eq:distortionmatrix} can be discretized by dividing
the interval between the observer and the source
into concentric $N$ spherical shells, each of comoving thickness $\Delta \chi$,
denoted by $\chi_k$ as the comoving distance to the middle of the
$k$-th shell and its related redshift $z_k$. 
A photon incoming on the $k$-th
shell at $\chi_k$ is deflected due to the presence of matter by an
angle $\boldsymbol\alpha^{(k)}$, which can be approximated by
\be
\boldsymbol\alpha^{(k)} = \frac{2}{c^2f_K(\chi_k)} \int^{\chi_k+\Delta \chi /2}
_{\chi_k-\Delta \chi /2} \nabla_{\beta} \Psi(\boldsymbol\beta(\boldsymbol\theta,\tilde{\chi}),\tilde{\chi})d
\tilde{\chi} = \nabla_{\beta} \Phi^{(k)}(\boldsymbol\beta(\boldsymbol\theta,\chi_k),\chi_k),
\ee
where we have defined the 2-D gravitational potential on the sphere as
\be
\Phi^{(k)}(\boldsymbol\beta(\boldsymbol\theta,\chi_k),\chi_k) = \frac{2}{c^2f_K(\chi_k)} \int^{\chi_k+\Delta \chi /2}
_{\chi_k-\Delta \chi /2} \Psi(\boldsymbol\beta(\boldsymbol\theta,\tilde{\chi}),\tilde{\chi})d
\tilde{\chi}.
\label{eq:phi2d}
\ee
Here, the notation $(\boldsymbol\beta(\boldsymbol\theta,\chi_k),\chi_k)$ means that the potential is
evaluated at the conformal look-back time $\chi_k$, when the
photon, coming from a source at distance $\chi_s$ from the observer at
position $\boldsymbol\theta$ on the celestial sphere, was at the position $\boldsymbol\beta^{(k)}$. 
The second derivatives can be combined into the
shear matrix ${\bf U}$:
\be
U^{(k)}_{ij} = \frac{\partial^2 \Psi^{(k)}(\boldsymbol\beta^{(k)}) }{\partial \beta^{(k)}_i
  \partial \beta^{(k)}_j} = \frac{\partial \alpha^{(k)}_i(\boldsymbol\beta^{(k)})}{\partial \beta^{(k)}_j},
\ee
being $\boldsymbol\alpha^{(k)}$ the lensing angle for the $k$-th shell. In case of the full-sky analysis the partial derivative operators have to be promoted to covariant derivatives \cite{Becker12}. 
The lensing potential for each matter shell $k$ is the solution of the Poisson equation, i.e.
\be
\nabla^2_{\hat{{\bf n}}}\Phi^{(k)}(\boldsymbol\beta^{(k)}) = 2 K^{(k)}(\boldsymbol\beta^{(k)}),
\label{eq:poiss}
\ee
where the convergence field $K^{(k)}$ at the $k$-th shell is
\be
K^{(k)}(\boldsymbol\beta^{(k)}) = \frac{4 \pi
  G}{c^2}\frac{D_A(\chi_k)}{(1+z_k)^2}\Delta^{(k)}_{\Sigma}(\boldsymbol\beta^{(k)}).
\label{eq:kappa1}
\ee
and $\Delta^{(k)}_{\Sigma}(\hat{{\bf n}})$ is the (projected)
surface mass overdensity, as in \cite{DasBode,Calabrese14}.
In Eq.~\eqref{eq:poiss} we dropped the term containing the
derivatives in the radial direction, ignoring thus 
long wavelength fluctuations along the line-of-sight via
the Limber approximation \cite{Jain00,Becker12}. As noted in \cite{Calabrese14}, the results of this approximation are particularly evident if we look at the angular power spectrum for the lensing potential of each single matter shell. However, if we look at
the overall effect after a sufficiently long photon path, the partial
derivatives in the transverse plane
commute with the integral evaluated along the whole line of sight, resulting in the cancellation of
line-of-sight modes as required in the Limber approximation of the
integral \cite{DasBode,Li2010,Becker12,Calabrese14}. The lensing potential on the sphere is related to $K^{(k)}$ via Eq.~\eqref{eq:poiss}, and
it can be easily computed by expanding both sides of the Poisson
equation in spherical harmonics. The quantity $K^{(k)}$ is directly computed when the matter distribution in
the shell is radially projected onto the spherical map, as explained in Sec.~\ref{sec:algorithm:mapmaking}.\\*
Given the deflection angle at each lens-plane, we can trace back the
light coming from a source at position $(\boldsymbol\theta,\chi_s)$ to the observer
after $N$ deflections:
\be
\boldsymbol\beta(\boldsymbol\theta,\chi_s) = \boldsymbol\theta -
\sum_{i=0}^{N-1}\frac{f_K(\chi_s -
  \chi_k)}{f_K(\chi_s)}\boldsymbol\alpha^{(k)}(\boldsymbol\beta^{(k)}).
\label{eq:fulllenseq_dis}
\ee
We can easily discretize the Eq.~\eqref{eq:distortionmatrix} as
\be
A^{N}_{ij}(\boldsymbol\theta, \chi_N) = \delta^K_{ij} - \sum^{N-1}_{k=0} \frac{D_{k,N}}{D_N}
U^{(k)}_{ip}(\boldsymbol\beta^{(k)},\chi_k)
A^{(k)}_{pj}(\boldsymbol\theta,\chi_k),
\label{eq:distortionmatrix_dis}
\ee
where we defined for simplicity $D_{k,N} = f_K(\chi_N - \chi_k)$, 
$D_k = f_K(\chi_k)$ while $N$ is the number of planes necessary to reach
the source at comoving distance $\chi_N$.  
The method in Eq.~\eqref{eq:distortionmatrix_dis} becomes computationally unfeasible very quickly, especially when we have a large number of lens planes covering a wide sky fraction. It, in fact, requires that for each $k\textrm{-th}$ iteration all the information of the $k-1$ deflections is kept. This becomes particularly problematic in the case of CMB where the source plane is located at very high redshift and at least 50 (or more) iterations are required to model the path of CMB photons. 
\cite{ValeWhite03,Hilbert09} proposed  a more efficient method 
that requires the combination of only two previous lens-planes
instead of the whole set as Eq.~\eqref{eq:distortionmatrix_dis}. The method has been validated on the full-sky in \cite{Becker12}. The
angular position $\boldsymbol\beta^{(k)}$ at the $k$-th shell is a
function of its two previous positions $\boldsymbol\beta^{(k-2)}$ and
$\boldsymbol\beta^{(k-1)}$ as 
\be
\boldsymbol\beta^{(k)}  = \left(1 -
\frac{D_{k-1}}{D_k}\frac{D_{k-2,k}}{D_{k-2,k-1}}\right) \boldsymbol\beta^{(k-2)}
 + \frac{D_{k-1}}{D_k}\frac{D_{k-2,k}}{D_{k-2,k-1}} \boldsymbol\beta^{(k-1)} 
 - \frac{D_{k-1,k}}{D_k} \boldsymbol\alpha^{(k-1)} (\boldsymbol\beta^{(k-1)}),
\label{eq:hilbert-beta}
\ee
and, by differentiating with respect to $\boldsymbol\theta$ as in
Eq.~\eqref{eq:distortionmatrix}, we obtain
the recurrence relation for the magnification matrix
\cite{Hilbert09,Becker12} as well:
\be
A^{(k)}_{ij} = \left( 1 -
\frac{D_{k-1}}{D_k}\frac{D_{k-2,k}}{D_{k-2,k-1}}\right) A^{(k-2)}_{ij} 
 + \frac{D_{k-1}}{D_k}\frac{D_{k-2,k}}{D_{k-2,k-1}} A^{(k-1)}_{ij}
 - \frac{D_{k-1,k}}{D_k} U^{(k-1)}_{ip} A^{(k-1)}_{pj}.\\
\label{eq:hilbert-a}
\ee
These relations require fewer arithmetic operations and memory usage
than the standard discretization of Eq.~\eqref{eq:distortionmatrix_dis}, therefore allowing us to compute
iteratively the magnification matrix for each light-rays from the
observer to any source. 
In the following we will also make use of the so-called magnification matrix in the first
order approximation \cite{Hilbert09} to assess the impact of the second-order effects and to distinguish those from numerical effects. In the multiple-lens formalism this takes the form 
\be
\label{eq:first_order_aij}
A^{(N), {\rm 1st}}_{ij}(\boldsymbol\theta, \chi_s) = \delta^K_{ij} - \sum^{N-1}_{k=0}
\frac{D_{k,N}}{D_N} U^{(k)}_{ij}(\boldsymbol\theta,\chi_k).
\ee

%% file: algorithm.tex

\section{The Algorithm} \label{sec:algorithm}
A detailed outline of the algorithm, including the construction of
the past light-cone and the map-making procedure, has been given in \cite{Calabrese14}.
In the following, we summarized the map-making procedure to produce lensing planes and the numerical and improvements specific to this work.

\subsection{N-body simulation}
The results reported in this were derived using the reference $\Lambda$CDM simulation belonging to the ``Dark Energy and Massive Neutrino Universe'' (DEMNUni) simulation project. We refer the reader to \citep{castorina2015,carbone2016} for a more detailed description of the simulations and for an extended discussion on the physical results issued by the project. The DEMNUni suite consists in a baseline $\Lambda$CDM model, and three modified $\Lambda$CDM
cosmologies characterized by three active neutrinos\footnote{The simulations do not account for an effective neutrino number $N_eff>3$. For example, neutrino isocurvature perturbations could produce larger $N_{eff}$ and thus affect galaxy and CMB power spectra \cite{boss}. However, these are excluded by present data (see e.g. \cite{divalentino}).} with different values of their total mass $\sum m_{\nu}$. All these simulations share the same total matter density $\Omega_m$, as
well as the same amplitude of primordial curvature perturbations and are based on a  {\it Planck-2013}  cosmology \cite{Planck2013_XVI}:
\begin{equation*}
\{ \Omega_{dm},  \Omega_{b}, \Omega_{\Lambda}, n_s, \sigma_8, H_0
  \} = \{ 0.27, 0.05, 0.68, 0.96, 0.83, 67 \: \rm{Km / s /
    Mpc} \}.
\end{equation*}
The employed $\Lambda$CDM simulation follows the evolution of $2048^3$ CDM particles in a cubic comoving volume $($2 $h^{-1} \text{Gpc})^3$ from redshift
 $z = 99$ to the present epoch (see e.g. Fig.~\ref{fig:kmax_nside}). {\bf Being $z=99$ the highest redshift probed by the simulation, the theoretical curves used for comparisons with numerical results in the following include contributions from matter perturbations $z\leq 99$. }
The mass resolution of the simulation at $z = 0$ is $M_{\rm CDM} = 8.27 \times 10^{10} M_{\odot} / h$ and the gravitational softening length is set to $\epsilon_s = 20$ kpc$/h$ corresponding to 0.04 times the mean linear inter-particle separation.
The simulation was carried out using a modified TreePM version of
 \lstinline!GADGET-3!\footnote{\url{http://www.mpa-garching.mpg.de/gadget}}
 \cite{Springel2005},
 specifically developed to include all the additional physical effects
 that characterize different neutrino models \cite{Viel2010}. This version of
 \lstinline!GADGET-3! follows the evolution of CDM and neutrino particles, treating them as two distinct sets of collisionless particles. Moreover, to test the effect the N-Body
resolution has on our results (see Sec.~\ref{subsec:nbody_res}), we considered a
Millennium-like numerical simulation, with the same background
cosmology and amplitude of primordial curvature perturbations as
described before. In this case the $2048^3$ CDM particles populate a
box with comoving side of 500 Mpc$/h$ and a mass resolution, for a DM particle at
$z=0$, of $M_{\rm CDM} = 8.27 \times 10^{8} M_{\odot} / h$.

\subsection{Map making}
\label{sec:algorithm:mapmaking}
Starting from different snapshots in time of the N-body simulation
we reconstructed the full-sky past lightcone of the observer back to the maximum redshift available in the simulation (in our case $z_{\rm max}=99$). Because the size of the simulation box is limited, we need to
replicate the box volume several times in space to fill the entire
observable volume between the observer and $z_{\rm max}$.
In order to avoid the repetition of the same structures along the line of sight and to mitigate the deficit of lensing power on angular scales comparable to the boxsize,
 we employed a specific randomization framework. We refer the reader to \cite{Springel01, Calabrese14} and references therein for a further discussion on this topic.
Following \cite{Carbone08, Carbone09, Calabrese14}, we divided the volume up to $z_{\rm max}$ into large spherical shells, each of
thickness comparable to the simulation boxsize.
All the simulation boxes falling into the same larger shell undergo
the same coherent randomization process, i.e. they are all
translated and rotated with the same random vectors generating an
homogeneous coordinate transformation throughout the 3D matter shell.  The coordinate transformation however changes from (large) shell to shell. We enforced the periodic
boundary conditions when performing box randomization to guarantee a smooth matter
distribution at the box edges and minimize numerical artifacts due to potential matter discontinuities at the boundaries.
The particles that fall within the radius of the
$k$-th spherical shell of width $\delta \chi$, ($\chi_k-\delta \chi/2,
\chi_k+\delta \chi/2$) are considered part of the shell, where $\delta\chi$ is the chosen average comoving thickness of the spherical shells. For our baseline setup $\delta\chi=150\textrm{Mpc}/h$. \\*
\noindent
Following the scheme proposed in \cite{DasBode, Calabrese14}, we then converted the position of a
particle of mass $m$ belonging to a given matter shell into its angular position on the 2D
sphere assuming a \lstinline!HEALPix! \footnote{\url{http://healpix.sourceforge.net}}
pixelization scheme. These maps are then projected into surface matter density maps $\Sigma^{\theta}$ assuming that in each pixel $p$ $\Sigma^{\theta}_p = \sum_{0}^{n} m_{i} / \Delta\Omega_{\rm pix}$, where $n$ is the number of particles per pixel and $\Delta\Omega_{\rm pix}$ its area in steradians.
For the results described in this paper we used maps having an angular resolution of 52 arcsec corresponding to a \lstinline!NSIDE! parameter $4096$.
\subsubsection{Systematics tests}\label{sec:maps-systematics}
\begin{figure}[!htbp]
\includegraphics[width=.5\textwidth]{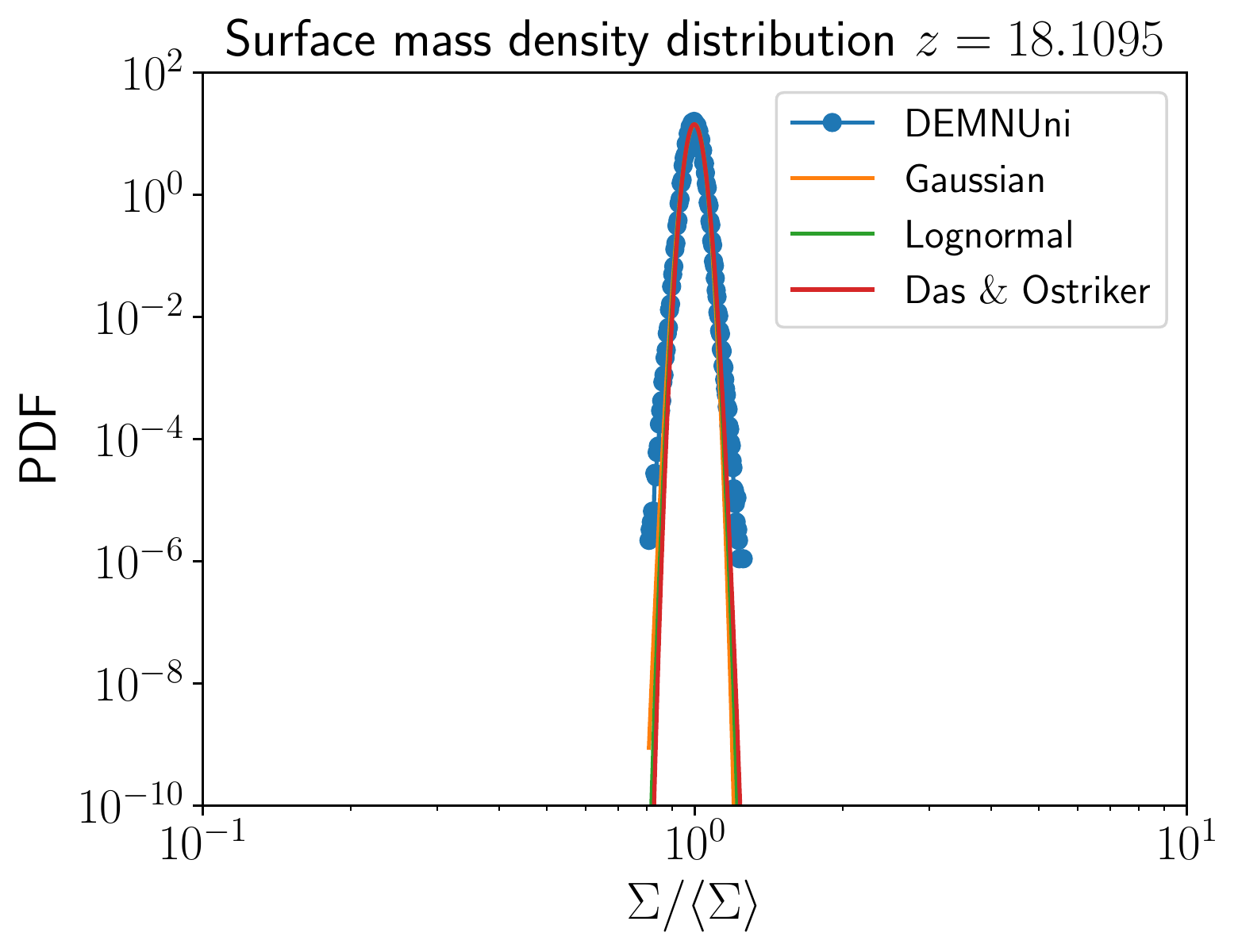}\includegraphics[width=.5\textwidth]{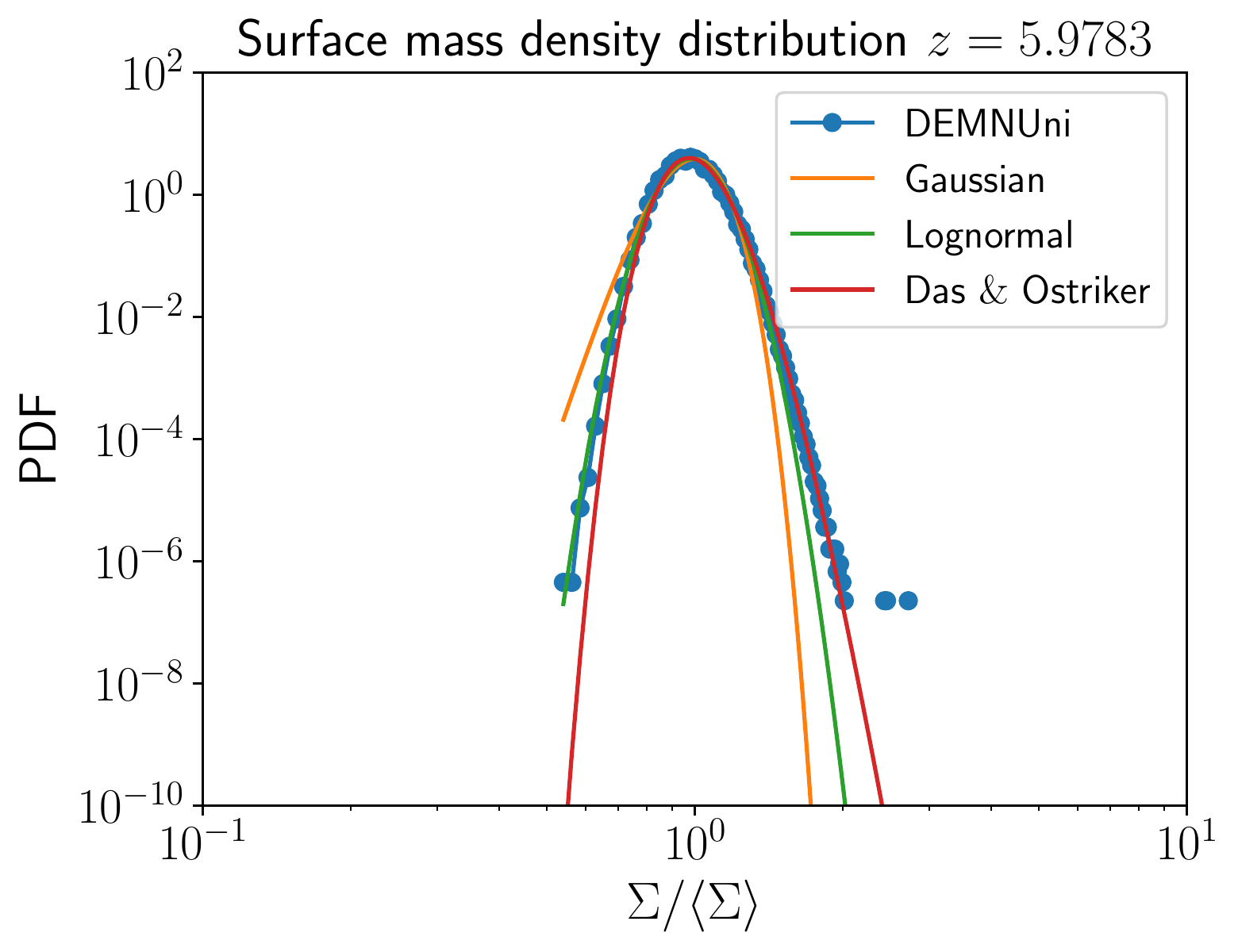}\\
\includegraphics[width=.5\textwidth]{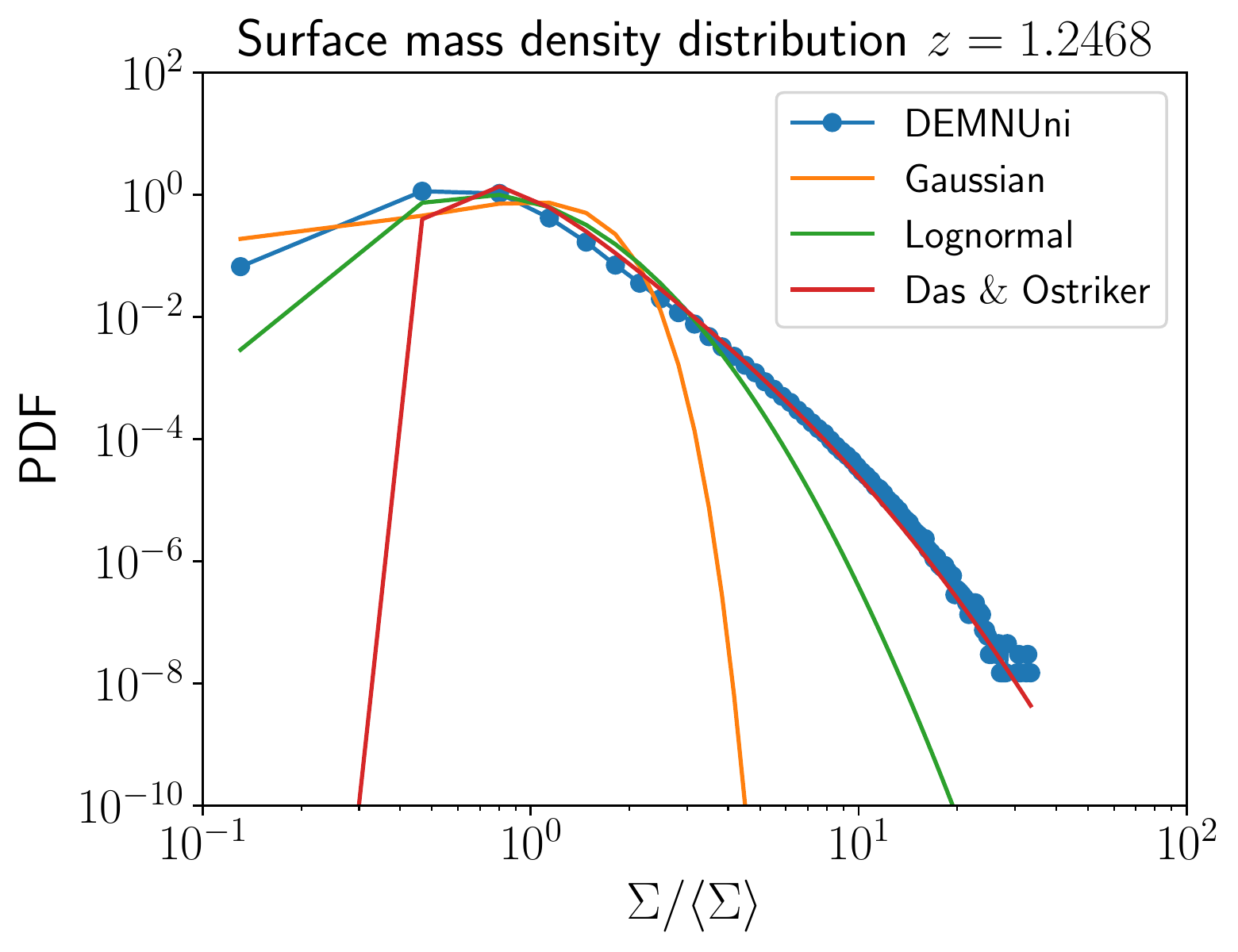}\includegraphics[width=.5\textwidth]{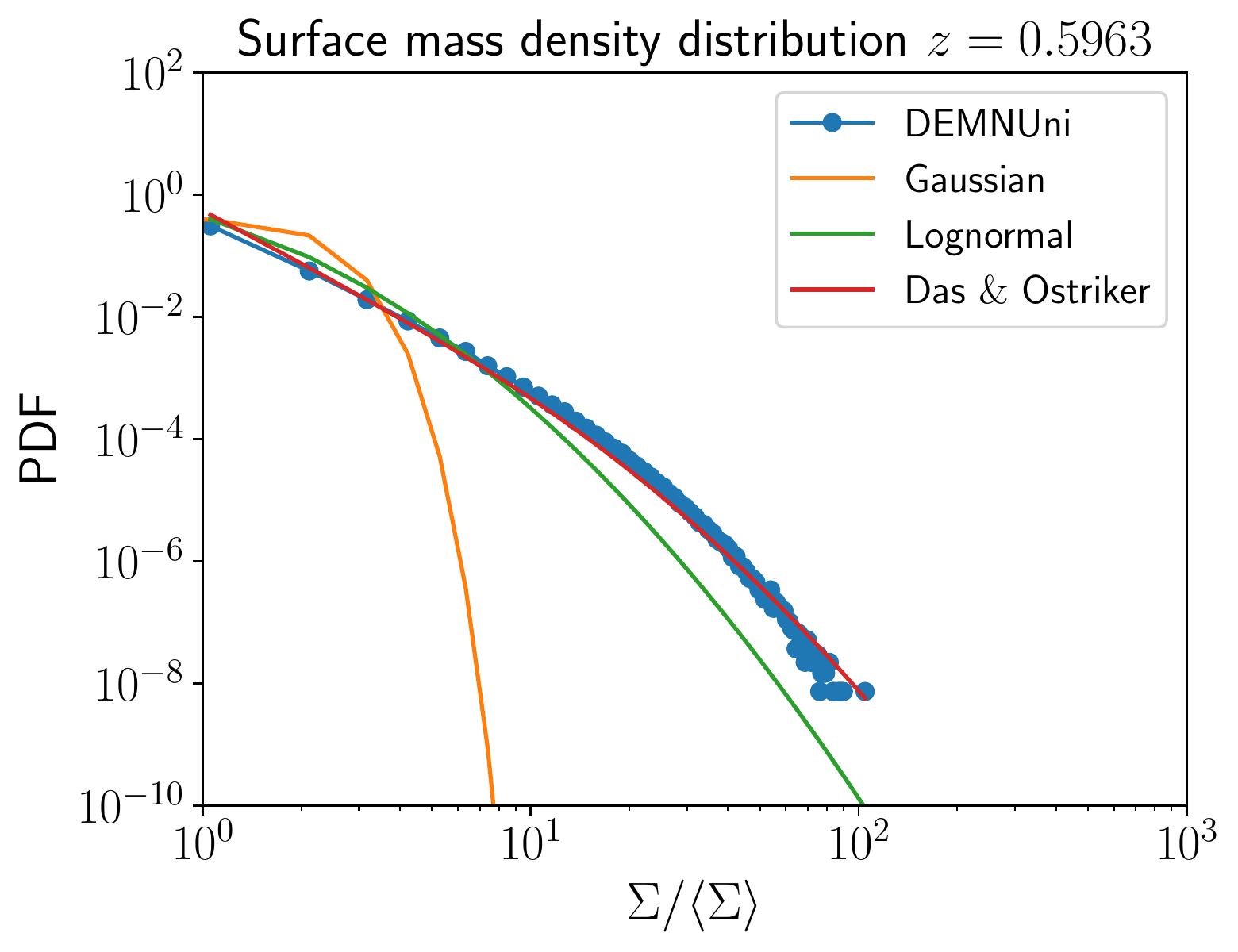}
\includegraphics[width=.5\textwidth]{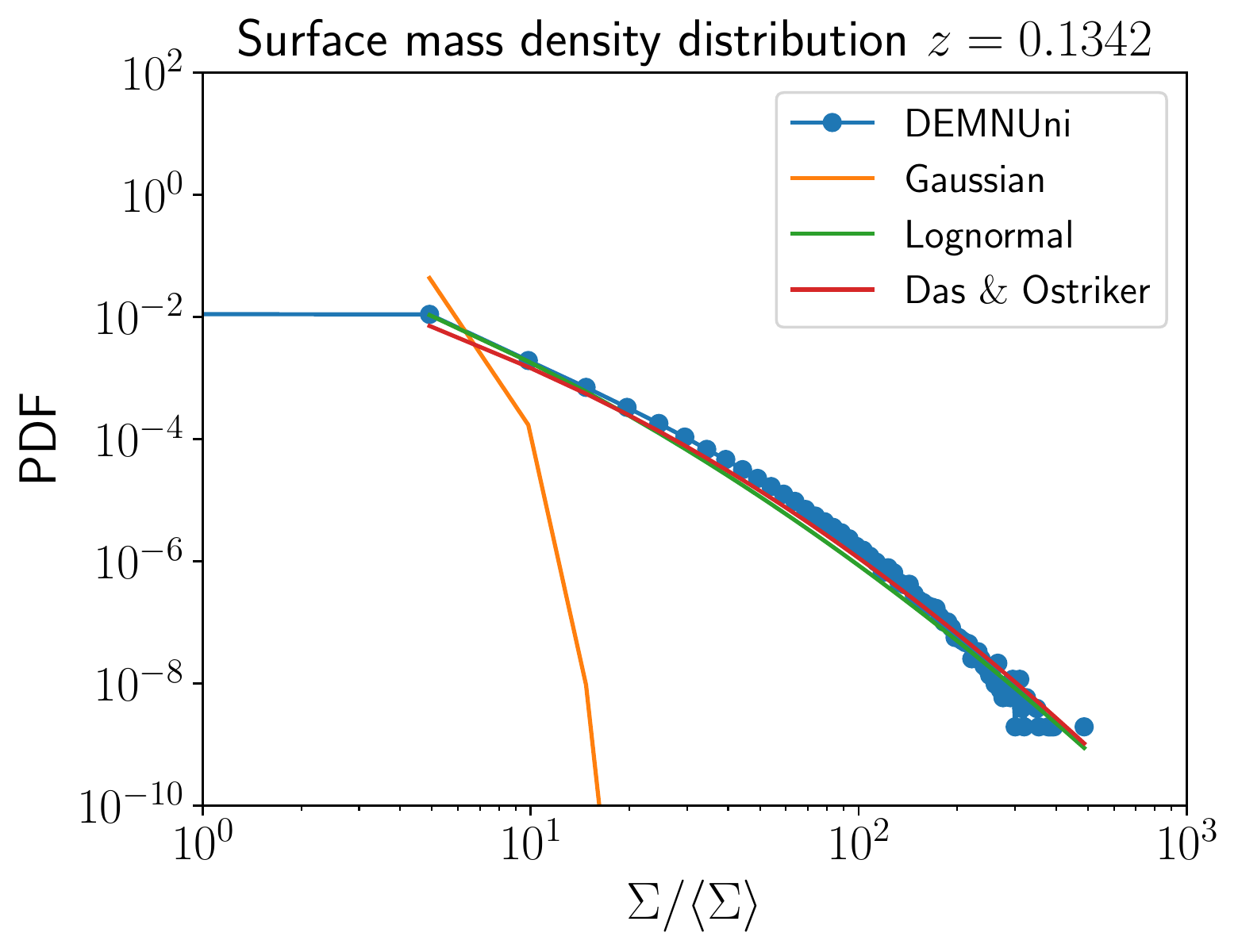}\includegraphics[width=.5\textwidth]{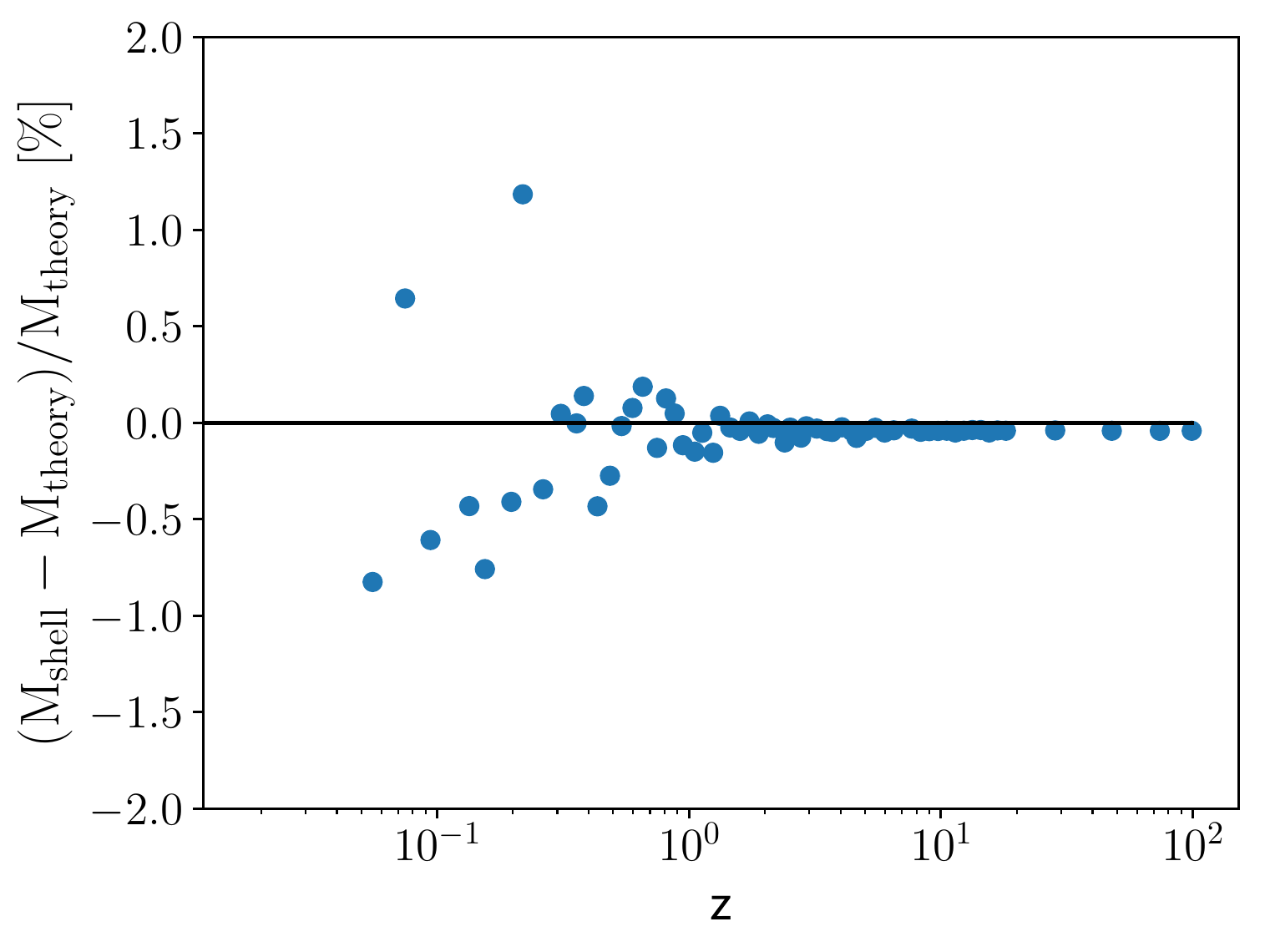}
\caption{PDF of the normalized surface mass density ($\Sigma/\langle\Sigma\rangle$) of the mass planes extracted from the DEMNUni simulations located at different redshift. The bottom right panel shows the fractional difference between the total mass in each mass plane extracted from the DEMNUni simulation and the theoretical value expected from the Planck 2013 cosmology.}
\label{fig:mass-histograms}
\end{figure}

Similarly to what was done in \citep{Calabrese14}, we performed several systematic test to assess the impact of possible systematic effect in the map-making procedure. The results of these tests are shown in Fig.~\ref{fig:mass-histograms}. We checked that the amount of mass present in each mass sheet was consistent with the amount of total mass expected from the cosmology and found differences of sub percent level on almost all shells and no greater than 2\% at the lowest redshifts bin.  We also checked the distribution of the surface mass density and found consistency with previous numerical work. The mass distribution is in fact well described by Gaussian or log-normal probability density function (PDF) at high redshift with a tail in the ultra-nonlinear regime well fitted by the phenomenological Das and Ostriker model \cite{DasOstriker} with smoothing scale set to the variance of each map.\\*
Once the surface mass density had been computed, we extracted the $k$-th shell surface mass density contrast  $\Delta^{(k)}_{
\Sigma^{\theta}}({\bf \hat{n}}) =  \Sigma^{\theta}({\bf \hat{n}}) - \langle \Sigma^{\theta}
\rangle$ and the convergence maps $K^{(k)}({\bf \hat{n}})$ (see Eq.~\eqref{eq:kappa1}) that will be used in the following.

\subsection{Raytracing and remapping}\label{sect:raytracing-remapping}
\begin{figure}[!htb]
\centering
\includegraphics[width=.8\textwidth]{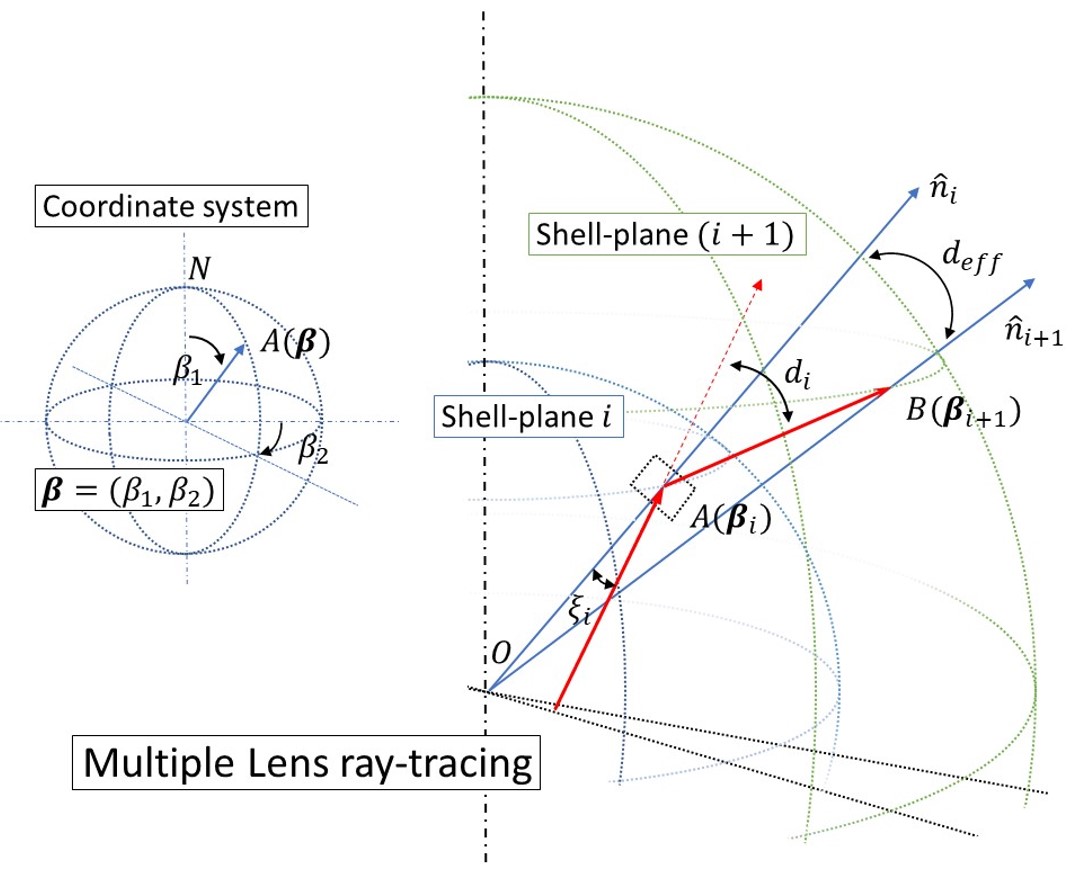}
\caption{Definition of the quantities employed in the multiple-lens raytracing algorithm.}
\label{fig:geometry}
\end{figure}

In order to propagate the CMB photons through the different shells we
used in \cite{Calabrese14} a
pixel-based approach exploiting the
publicly available code \lstinline!LensPix! \cite{Lewis05}. For this
work, we changed this part of the algorithm and adopted a modified version of the
\lenstwo\ code \cite{Fabbian13}. In particular we implemented the raytracing algorithm of Eq.~\eqref{eq:hilbert-beta} and Eq.~\eqref{eq:hilbert-a}
while carefully preserving the optimal data and workload distribution of the original algorithm. \\*
The algorithm operates as follow. Starting from
the convergence harmonic coefficients $K^{(k)}_{\ell m}$ it extracts the lensing potential of each shell solving the Poisson equation in the harmonic domain  as $\psi^{(k)}_{\ell m}=-2K^{(k)}_{\ell m}/[\ell(\ell+1)]$. It then computes the deflection field for each lensing plane $\boldsymbol\alpha^{(k)}$ assuming the deflection field
as a purely gradient field (i.e. a spin-1 curl-free vector field) having E and B decomposition
\be
_{1}\alpha^{E}_{\ell m}{}^{(k)}=\sqrt{\ell(\ell+1)}\psi^{(k)}_{\ell m} \qquad _{1}\alpha^{B}_{\ell m}{}^{(k)}=0.
\label{eq:deflection-field-calculation}
\ee
Note that the lensing potential $\psi$ is simply the 2-D gravitational potential of Eq~\eqref{eq:phi2d} multiplied by a geometrical lensing weight $f_K(\chi_s - \chi_k)/f_K(\chi_s)$. Once the deflection field is obtained, it remaps the CMB field as
\be
\boldsymbol\beta^{(k)} = \boldsymbol\beta^{(k-1)} + \boldsymbol d^{(k)}
\label{eq:raytracing}
\ee
where $ \boldsymbol\beta^{(k-1)}$ represents the position of an incoming
CMB photons on the $k$-th lensing plane and ${\bf d}^{(k)}$ the total deflection field at each $k$-th lensing plane. This takes into account the deflection field $\boldsymbol\alpha^{(k)}$ and the incoming direction of the photon with respect to the normal to the surface $\boldsymbol \xi^{(k)}$ so that ${\bf d}^{(k)}=\boldsymbol\alpha^{(k)}+\boldsymbol\xi^{(k)}$. In the code, we adopted the algorithm of \cite{DasBode} to compute $\boldsymbol\xi^{(k)}$, which can be shown to be equivalent to the recurrence equation of Eq.~\eqref{eq:hilbert-beta}\footnote{We renamed  $\boldsymbol\xi$ the $\boldsymbol\beta$ of \cite{DasBode} to avoid confusion with the notation adopted in the text so far.}.  In the case $k=1$, $ \boldsymbol\beta^{(0)}= \boldsymbol\theta$ i.e. the unlensed CMB photon position. Following \cite{Becker12} we accounted for the change of the local tensor basis when parallel transporting the lensing jacobian along the geodesic connecting the old and displaced position at each deflection plane adopting the formalism of \cite{challinor-chon}.  For sake of clarity we show in Fig.~\ref{fig:geometry} the definition of the angles adopted in the algorithm. 
Although algorithmically similar to \lenspix,
\lenstwo\ offers several advantages when performing full-sky raytracing simulation for application that require the use of the highest possible angular resolution.
The advantages lie mainly in the nearly perfect scalability and the very low memory overhead inherited from the \stwo\ library\footnote{\url{http://www.apc.univ-paris7.fr/APC_CS/Recherche/Adamis/MIDAS09/software/s2hat/s2hat.htm}} \citep{hupca2011, szydlarski2011, fabbian2012}. These allow to manipulate maps and compute spherical harmonics at very high resolution required by this specific application. The code is also very flexible as it works for multiple pixelization schemes beyond the commonly used \healpix. In particular, for this work, we resampled the displacement field and performed the raytracing part on grids based on the Equidistant Cylindrical Projection (ECP) pixelization \citep{muciaccia1997}. Thanks to its symmetry properties, this pixelization supports fast spherical harmonic transforms and nearly perfect quadrature for band-limited functions, minimizing thus aliasing effects that may become important in codes of this type.\\*
\lenstwo\ implements the raytracing using a simple yet powerful pixel-based lensing
method adopting an efficient Nearest Grid Points (NGP) assignment scheme to evaluate the
source plane (or, in this case the lensing plane) along the displaced direction. We thus assign to
every deflected direction a value of the sky signal computed at the
nearest grid points defined by the centers of a pixel of the assumed
pixelization scheme. The NGP assignment is quick and allows to control the accuracy of the method and the signal smoothing scale through the parameters defining the
grid resolution alone. Interpolation methods on the sphere, though helpful in cutting numerical costs, might modify the underlying signal and alter its statistical properties especially at the smallest scales, where we expect the effects of our interest to lie. We refer the reader to \citep{Fabbian13} for a further discussion on this topic together with an accurate benchmark of \lenstwo.

\subsection{Derivatives computation on the sphere}
In order to propagate the lensing jacobian together with the displaced photon direction it is necessary to compute second derivatives together with first derivatives of the projected gravitational potential (see Eq.~\ref{eq:distortionmatrix_dis},~\ref{eq:fulllenseq_dis}). It is possible to compute efficiently the derivatives of this field on the sphere using the routine \lstinline!alm2mapspin_derv! included in the \stwo\ library. This routine can compute the first partial derivatives of an arbitrary spin-$s$ field on the sphere, $\partial_{\theta},\partial_{\phi}$, with respect to the spherical coordinates ($\theta,\phi$). The routine implements an efficient recurrence relation for the Legendre polynomials part of the harmonics and for their derivatives starting from their expression in terms of Wigner D-matrices \cite{Varshalovich}. This recurrence is similar to the one implemented in the \lstinline!HEALPix!
library, where however only the spin-2 field derivatives are available. Numerical tests conducted so far have
shown that derivatives of the same field on a sphere, computed with \stwo\
or \lstinline!HEALPix!, are equal within the numerical error.
Thanks to the capability of the routine to compute derivative for arbitrary spin-$s$ field it is possible to compute the second derivatives of the matter distribution at the same time when computing the deflection field (see Eq.~\eqref{eq:deflection-field-calculation}). As shown in Appendix~\ref{sec:app_derivative_spin} in fact, the second derivatives of the lensing potential can be naturally related to the first derivatives of the spin-1 deflection field. Thus, the deflection field and the second derivative of the lensing potential can be obtained with a single call to the spin-$1$ Legendre polynomial recurrence. This minimizes the extra computational overhead required to compute the second derivatives given that the Legendre polynomial recurrence is the heaviest part of each spherical harmonic transform algorithm (see, e.g., \cite{szydlarski2011, fabbian2012}). Once partial derivatives are available it is then straightforward to extract the second covariant derivatives, which are the key quantity for a full-sky formalism adopted in this work.

%% file: results.tex
\section{Simulation results}
\label{sec:results}
\subsection{Lensing observables angular power spectra}\label{subsec:magnif_spectra}
\begin{figure}[!htb]
\centering
\includegraphics[width=.495\textwidth]{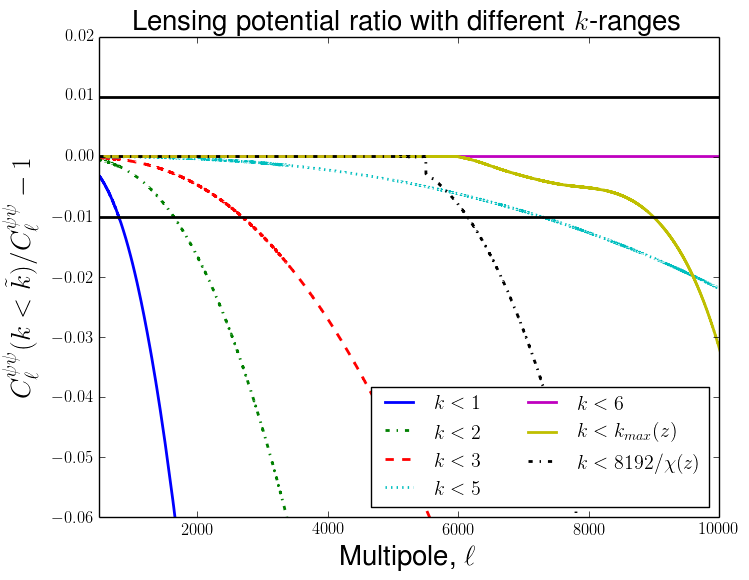}
\includegraphics[width=.49\textwidth]{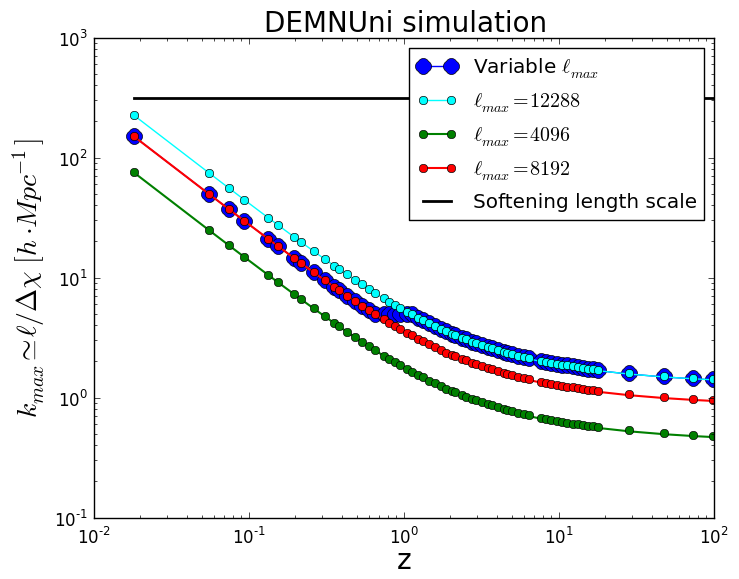}
\caption{\emph{Left}: fractional difference between the lensing
 potential angular power spectrum as recovered by
 \lstinline!CAMB! using modes up to $k=100$ and the same signal computed with a reduced wavenumber range $k<k_{\rm max}$. Black horizontal lines show the 1\%
 precision threshold. \emph{Right}: maximum wavenumber $k_{\rm max} \approx \ell_{\rm max} / \chi(z)$ included in the simulation as a function of redshift for different value of band-limit. The variable $\ell_{\rm max}$ case is the reference option used in this work. The softening length scale of our N-Body simulation is shown as a black solid line.}
\label{fig:kmax_nside}
\end{figure}

The first problem that we addressed using our raytracing algorithm was the evaluation of the full set of components of the CMB lensing jacobian. We thus set the position of our source at the CMB last scattering surface ($z^{*}\approx 1100$) and propagated the light ray position to redshift $z=0$. For these runs we assumed that no lensing took place between the maximum redshift covered by the DEMNUni simulation and the last scattering surface. 
One of the free parameters of the algorithm, together with the spatial resolution of the simulation, is the band-limit value $\ell_{\rm max}$ adopted to solve the Poisson equation and extract the lensing potential $\psi^{(k)}_{\ell m}$ from each convergence map $K^{(k)}$ (see Sec. \ref{sect:raytracing-remapping}). The choice of this band-limit defines effectively the Fourier modes $\Psi(k)$ of the matter distribution that are included in the simulation and thus can affect the precision of the results itself.\\* 
We decided to adopt a pragmatic approach to identify a suitable choice of band-limit for this application and proceeded as follows. We first computed the expected theoretical CMB lensing power spectrum for the Planck 2013 cosmology with \lstinline!CAMB! and then evaluated the error introduced with respect to this reference quantity when the same calculation was performed including only the modes $k<k_{\rm ref}$ of the matter power spectrum. We report the results of this calculations for reference in Fig.~\ref{fig:kmax_nside}. We set our goal to be the recovery of the effective CMB lensing potential with roughly 1\% precision up to $\ell\approx 8000$, thus we had to include modes of the matter distribution up to $k_{\rm max}\approx 5 h/Mpc$. Following this reasoning, we then translated this constraint in the value of the band-limit parameter to be used at a given i-th shell using the Limber approximation, i.e. $\ell^{(i)}_{\rm max}=\ell_{\rm max}^{(i), \rm theory}\equiv k_{\rm max}\chi_{i}$. However, it turned out that for several shells $\ell_{\rm max}^{(i)}$ became higher than the band-limit supported by the \healpix\ grid where we sampled the $K^{(k)}$ field ($\ell_{\rm max}^{grid}=3 \cdot$\nside). For these cases we set $\ell_{\rm max}^{(i)}=\ell_{\rm max}^{grid}$ while we chose to fix $\ell_{\rm max}^{(i)}$ to a conservative reference value of 8192 if $\ell_{\rm max}^{(i),\rm theory}<8192$. In Fig.~\ref{fig:kmax_nside} we show the modes of the matter distribution effectively included in the raytracing simulation as a function of redshift given the choice of $\ell_{\rm max}^{(i)}$ just presented. In Fig.~\ref{fig:kmax_nside} we also report the corresponding precision on the lensing potential obtained with the same choice of band-limit. As it can be seen from these figures, the optimized procedure improves the achievable precision at $\ell=8000$ from the 6\% obtained assuming the same band-limit $\ell_{\rm max}=8192$ for all shells to the targeted 1\% level. We, in fact, cannot resolve modes $k<5$ at all redshift because we are limited by the band-limit imposed by the resolution of our convergence maps and the optimized band-limit mitigates the impact of lack of power on those scales. This problem could be bypassed either fixing a band-limit to higher value (increasing the computational cost) or producing convergence maps at higher \healpix\ resolution. The latter option would in fact allow to solve the Poisson equation at higher multipoles. As it will be shown in the next sections, however, we found no indication that this baseline setup suffers from a lack of power at small angular scales, and therefore we decided not to pursue this option due to its higher computational cost. 
\noindent

\begin{figure}[!htbp]
\centering
\includegraphics[width=.5\textwidth]{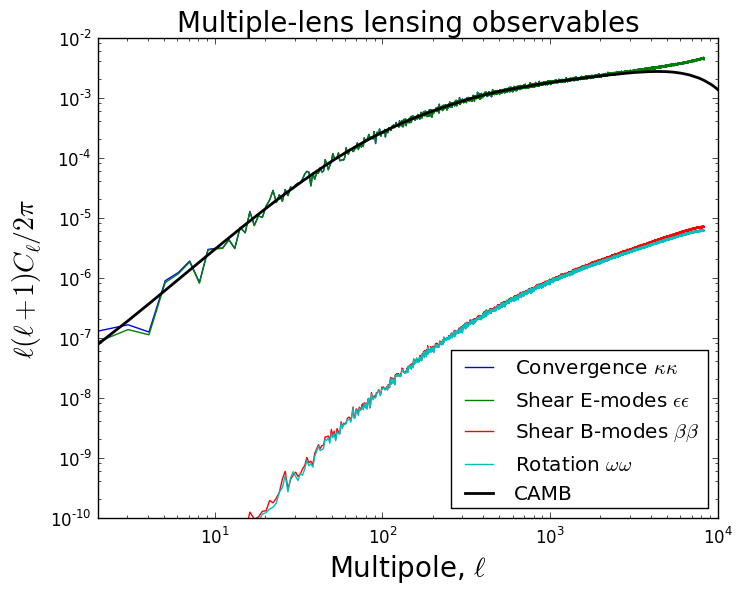}\includegraphics[width=.5\textwidth]{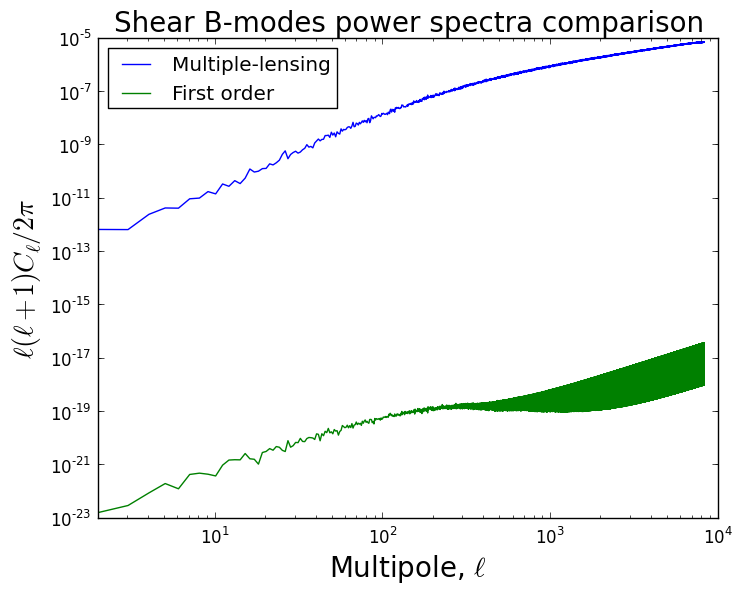}
\caption{\emph{Left}: angular power
 spectra for CMB lensing observables $\kappa\kappa$ (blue),
 $\epsilon\epsilon$ (green), $\omega\omega$ (cyan)
 and $\beta\beta$ (red). The reference convergence angular power spectrum as computed by \camb\ is shown in black. \emph{Right}: Born (green) and beyond-Born (blue) shear B-modes derived from the DEMNUni simulations.}
\label{fig:magnification_spectra}
\end{figure}

Once $\psi^{(k)}_{\ell m}$ had been obtained, we converted them into Fourier modes of the deflection field and resampled the deflection field and all its second derivatives required by the algorithm on a very high resolution ECP grid on which we later performed the raytracing step. 
We note that this part of the algorithm is the most computationally expensive step and therefore a more careful optimization of the choice of the band-limit parameter according to the users needs is crucial to reduce the computational cost of these kind of simulations.\\*
In the following sections we will show results derived employing ECP grids characterized by the same number of isolatitude rings on the sphere $N_{\theta}$ and number of pixel per isolatitude ring $N_{\phi}$. The map resolution is therefore non uniform across the sky (with smaller pixels close to the poles and bigger close to the equator) but the area of the coarser pixels is well approximated by $\theta_{\rm res} = \sqrt{\frac{4\pi}{N_{\rm pix}}}$. We thus express the map resolution in terms of the parameter $\sqrt{N_{\rm pix}}=\sqrt{N_{\theta}\cdot N_{\phi}}$.
In Fig.~\ref{fig:magnification_spectra} we show the angular power
spectrum of the different lensing observables extracted from the CMB lensing jacobian at $z=0$ -- convergence,
$C^{\kappa\kappa}_{\ell}$, rotation $C^{\omega\omega}_{\ell}$ and E-
and B-modes of lensing shear, $C^{\epsilon\epsilon}_{\ell}$,
$C^{\beta\beta}_{\ell}$. For these simulations we employed an ECP grid with $\sqrt{N_{\rm pix}} = 262144$ which corresponds to $\theta_{\rm res} \approx 3^{\prime\prime}$ to derive these results. The previous set of parameters represents
the highest setup in terms of resolution and band-limit shown in
this paper, although we tested that setups with $\sqrt{N_{\rm pix}} = 524288$ are achievable with the current numerical implementation of the code.\\*
The simulated convergence and E-modes power spectra are in good agreement
with the HALOFIT based results derived with \camb. For the purpose of comparing the two quantities we restricted the integration of the matter evolution in the redshift range covered by the N-body simulation. The accuracy of the results is compatible with the uncertainties proper of the HALOFIT fitting formulae \citep{Takahashi12}. Matter perturbation on scales larger than the box-size of the simulation used to construct the lightcone should only be partially recovered by the stacking procedure employed in the map-making step \citep{Calabrese14}. However simulations having a box-size of 2Gpc like the ones of the DEMNUni suite seem to reproduce well the convergence power spectrum at large angular scales for all practical purposes.\\* 
In Fig.~\ref{fig:magnification_spectra} we can also observe a non-zero image rotation $C^{\omega\omega}_{\ell}$ and shear B-modes $C^{\beta\beta}_{\ell}$ power spectrum which are due to multiple deflections of photons and are a pure imprint of beyond-Born corrections properly resolved by our simulations. We cross-checked this fact performing the raytracing with the same numerical setup but propagating the CMB lensing jacobian at first odder, i.e. in the Born approximation (Eq.~\eqref{eq:first_order_aij}). The shear B-modes extracted from the first-order simulations are shown in Fig.~\ref{fig:magnification_spectra} together with the beyond-Born result and they are consistent with numerical noise. We note that the second mixed covariant derivatives of a scalar field on the sphere commute, therefore we did not show the image rotation at first-order because it is identically zero by construction. All these results answer and confute the statement made in \cite{Hagstotz14} according to which the multiple lens approach is not able to properly resolve beyond-Born corrections.

\noindent

\subsection{Consistency tests}\label{sec:consistency}
\begin{figure}[!htbp]
\centering
\includegraphics[width=.85\textwidth]{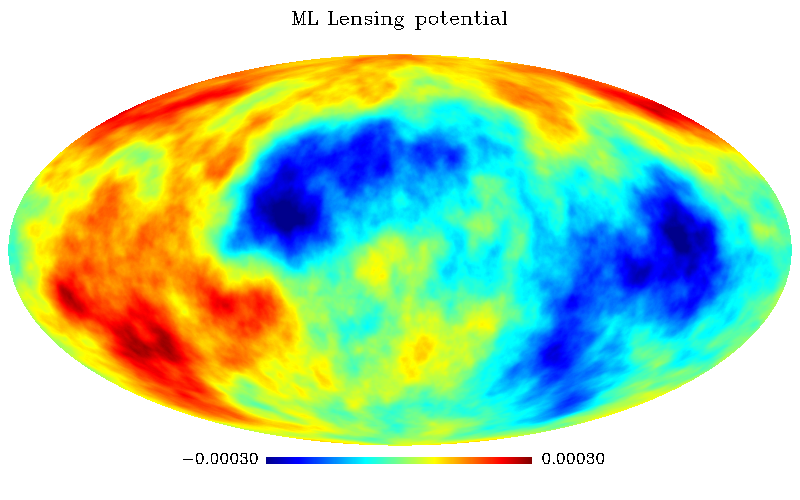} \\\includegraphics[width=.85\textwidth]{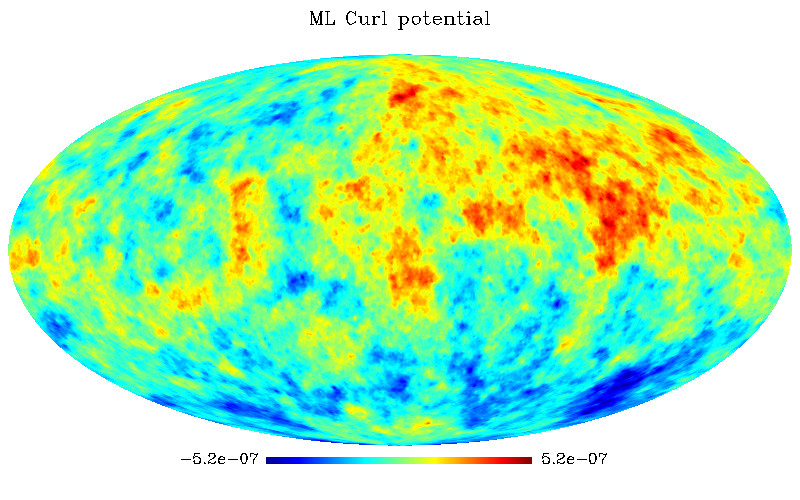} 
\caption{Maps of the effective CMB lensing potential $\psi^{\rm eff}$ (top) and curl potential $\Omega^{\rm eff}$ (bottom). The potentials were extracted with the ML approach from the effective displacement of CMB photons in the DEMNUni simulation.}
\label{fig:scalar-potentials}
\end{figure}

We assessed the consistency of our results and the potential presence of numerical artifacts in two ways. First, we tested if the consistency relations between the components of the lensing jacobian were satisfied (see Appendix~\ref{sec:appendix}). In addition, we tested if these also satisfied the consistency relation that can be derived between the E and B-modes of the effective displacement field of the rays. Following the notation in Eq.~\eqref{eq:raytracing} we can define the effective displacement field ${\bf d}^{\rm eff}$ after $N$ deflections as
\be
 \boldsymbol\beta^{N} = \boldsymbol\theta - {\bf d}^{\rm eff}
\ee
where $ \boldsymbol\theta$ is the photon un-displaced direction. Because we know the initial and final position of the simulated light rays, it is possible to invert the equation above using spherical triangles identities \cite{Lewis05} and derive ${\bf d}^{\rm eff}$. We can extract the effective potentials of \ref{eq:hs03} from the E/B decomposition of this vector (i.e. spin-1) field as
\begin{equation}
{\bf d}^{\rm eff,E}_{\ell m} = \sqrt{\ell(\ell+1)}\psi^{\rm eff}_{\ell m} \qquad {\bf d}^{\rm eff,B}_{\ell m} = \sqrt{\ell(\ell+1)}\Omega^{\rm eff}_{\ell m},
\end{equation}
In Fig.~\ref{fig:scalar-potentials} we show the maps of the effective potentials extracted from our simulations\footnote{From now on we will drop the \emph{eff} superscript for the effective scalar potentials for sake of brevity.}. 
 
\noindent

\begin{figure}[!htbp]
\centering
\includegraphics[width=.32\textwidth]{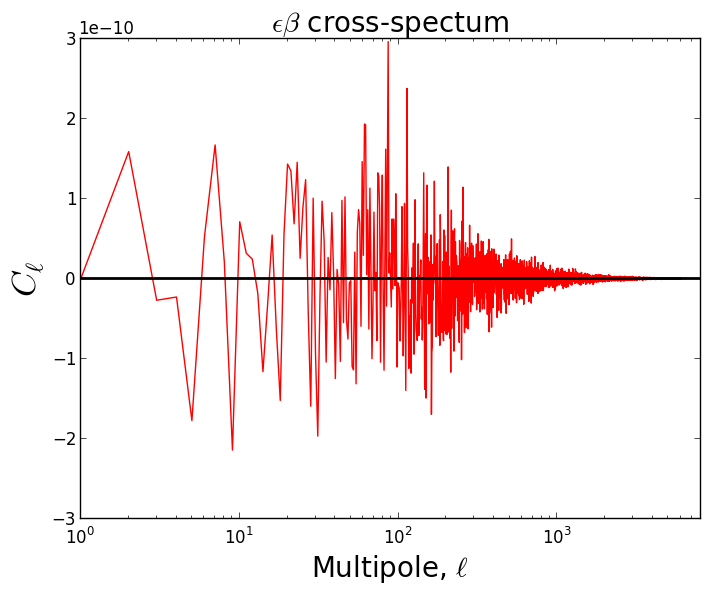}
\includegraphics[width=.32\textwidth]{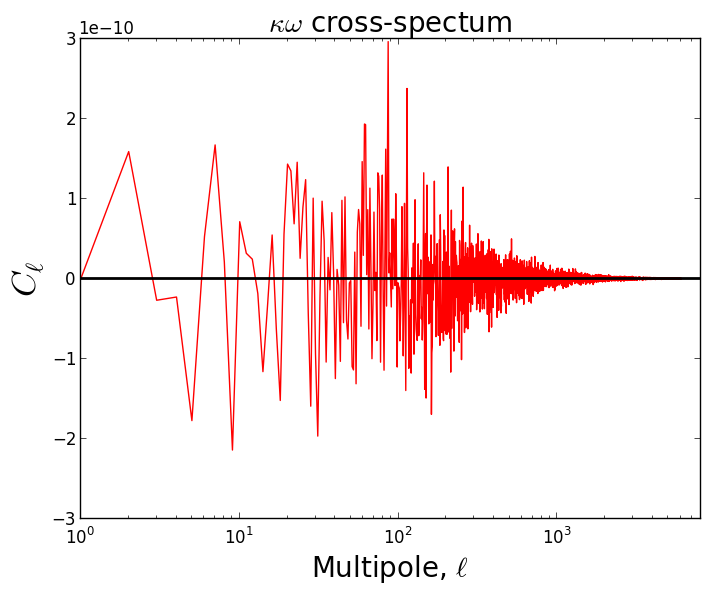}
\includegraphics[width=.32\textwidth]{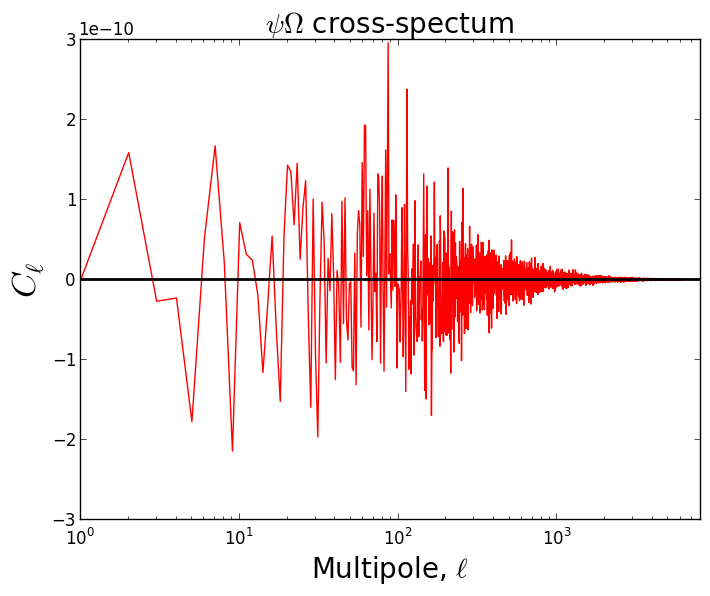}
\caption{Angular cross-spectra for different lensing observables
 combinations. Black horizontal shows the zero line reference. 
}
\label{fig:consistency_cross_spectra}
\end{figure}

\begin{figure}[!htbp]
\centering
\includegraphics[width=.5\textwidth]{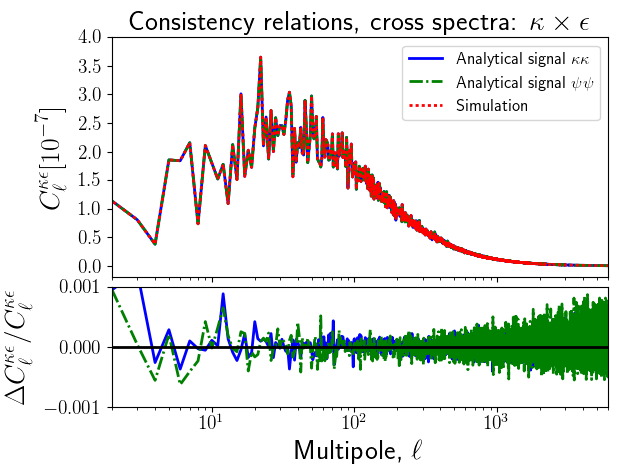}
\includegraphics[width=.488\textwidth]{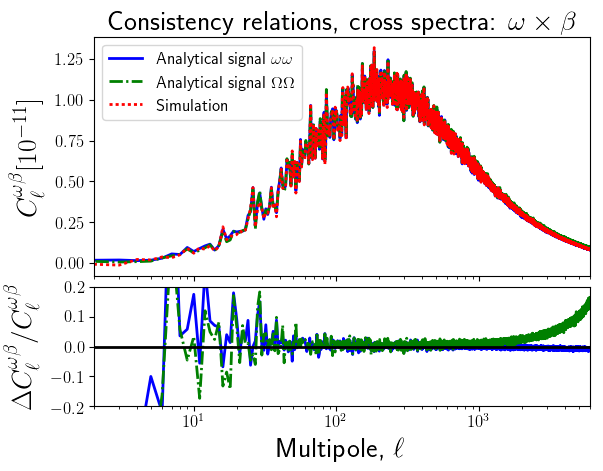}
\caption{\emph{Top left}: Angular cross-power spectra $\kappa\epsilon$
 constructed by analytical expression compared with the one extracted
 from simulations of magnification
 matrix $\kappa \times \epsilon$ (red line). Green and blue line represent angular cross-
 spectra constructed from power spectra of $\kappa\kappa$ and lensing
 potential $\psi\psi$, respectively. \emph{Top right}:
 angular cross-power spectrum for the rotation fields
 $\omega\beta$. \emph{Bottom}: fractional difference between the analytical constructions and the recovered cross spectra.
}
\label{fig:consistency}
\end{figure}

The second test we performed was computing
cross power spectra of lensing observables that are supposed to be null in absence of systematics (see Appendix~\ref{sec:appendix}). In particular we
looked at combinations of $\kappa \times \beta$, $\epsilon \times
\beta$, $\omega \times \epsilon$, $\kappa \times \omega$ for the
magnification matrix, and $\psi \times \Omega$ for the lensing potentials
fields themselves. 
In Fig.~\ref{fig:consistency_cross_spectra} we show a selection of these
cross-spectra omitting the ones that can be derived using the lensing consistency relations. All are clearly consistent with a null spectrum with no significant trend at all scales. 
\noindent
In order to test the consistency relations between lensing observables we adopted a cross-spectra technique and focused on cross-spectra combinations $\kappa \times \epsilon$ and $\omega \times \beta$. As
it can be derived using the formulae in Appendix~\ref{sec:appendix}, these cross-spectra can be
constructed from the angular power spectra of the lensing fields
themselves, as
\begin{eqnarray}\label{eq:kappa-epsilon}
C^{\kappa\epsilon}_{\ell} &=& C^{\kappa\kappa}_{\ell}
\sqrt{\frac{(\ell+2)(\ell-1)}{\ell (\ell+1)}}, \\
C^{\omega\beta}_{\ell} &=& C^{\omega\omega}_{\ell}
\sqrt{\frac{(\ell+2)(\ell-1)}{\ell(\ell+1)}},
\end{eqnarray}
where the angular power spectra on the r.h.s. of the equations is
computed directly from simulated maps. A similar construction can be obtained instead using the effective lensing potentials, as
\begin{eqnarray}\label{eq:kappa-epsilon-disp}
C^{\kappa\epsilon}_{\ell} &=&
\frac{1}{4}(\ell+1)\ell\sqrt{(\ell+2)(\ell+1)\ell(\ell-1)} C^{\psi\psi}_{\ell},\\ \nonumber
C^{\omega\beta}_{\ell} &=&
\frac{1}{4}(\ell+1)\ell\sqrt{(\ell+2)(\ell+1)\ell(\ell-1)} C^{\Omega\Omega}_{\ell}.
\end{eqnarray}
In Fig.~\ref{fig:consistency}, we compare cross-spectra as
extracted from simulations with the two
analytical constructions of Eqs.~\eqref{eq:kappa-epsilon},\eqref{eq:kappa-epsilon-disp}. We can see that the consistency relation involving the gradient-like components is recovered with exquisite accuracy (below 0.1\%) for both $\kappa$ and $\epsilon$ and $\epsilon$ and $\psi$. Conversely, the consistency relation between the shear B-modes and the rotation 
field does not reach the same level of accuracy, nevertheless it is satisfied at the 1\% level at small angular scale and is better than numerical results presented so far in the literature (see e.g. \cite{Becker12}). The consistency relation derived using the $\Omega$ effective potential is satisfied at the 10\% level and we can see a clear excess of power at small-scales ($\ell>1000$). As discussed in the following, the stability of the results of the ${\bf d}^{\rm eff,B}$ and shear B-modes with respect to the choice of numerical setup gives an indication that the cause of this discrepancy is likely purely numerical. A
residual of aliasing effect might in fact be present in the shear and
${\bf d}^{\rm eff,B}$ affecting thus the consistency
relation involving these two quantities. As explained in
\cite{Fabbian13}, the B-modes of spin fields are much more
sensitive to numerical effects than the scalar quantities, and controlling residual
resolution effects may become crucial.

\subsection{Testing resolution effects}\label{subsec:stability}
\begin{figure}[!htbp]
\centering
\includegraphics[width=.48\textwidth]{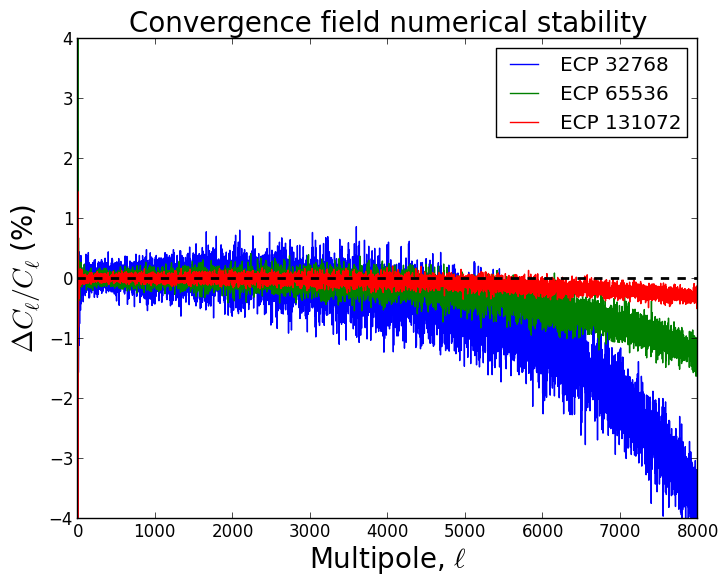}
\includegraphics[width=.48\textwidth]{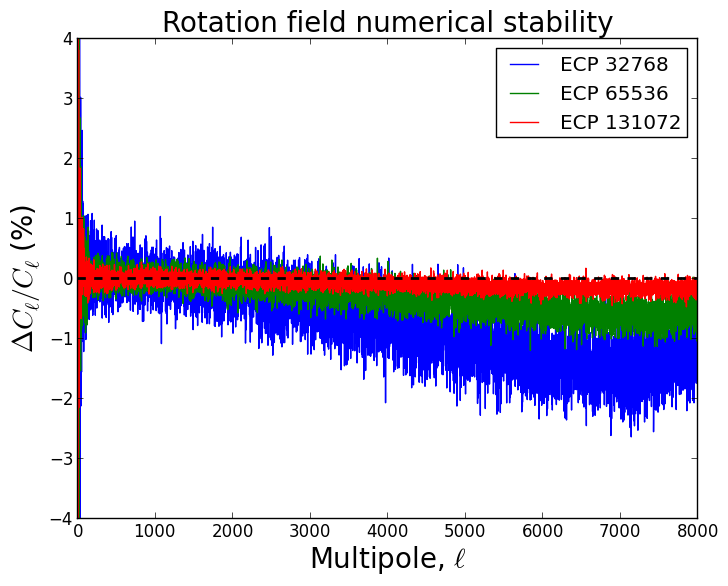}
\caption{\emph{Left}: Fractional difference of convergence angular power spectrum, $C^{\kappa\kappa}_{\ell}$ obtained fixing the band-limit and varying the spatial resolution of the raytracing from $\sqrt{N_{\rm pix}^{(i)}}$ and $\sqrt{N_{\rm pix}^{(i+1)}} $where $\sqrt{N_{\rm pix}^{(i)}} \in \{32768, 65536, 131072, 262144\}$.
\emph{Right}: same as the right power but for the rotation power spectrum. $C^{\omega\omega}_{\ell}$}
\label{fig:resol_k_omega}
\end{figure}

We additionally tested the stability of the simulated $\kappa, \gamma$ and $\omega$ maps with respect to different numerical parameters used for the raytracing. In Fig.~\ref{fig:resol_k_omega} we show the effect of the map
resolution used in the raytracing on the reconstructed angular power spectra of the lensing observables. The panels
show the fractional difference between results of raytracing performed at progressively higher spatial resolution but keeping the same band-limits choice discussed in Sec.~\ref{subsec:magnif_spectra}. 
For the kind of resolution considered for this test we recovered a percent level precision on $\kappa$ and $\omega$ already using a resolution of $\sqrt{N_{\rm pix}}=65538$ at scales $\ell\approx 5000$. This level of precision reaches the sub-percent level on $\ell<8000$ using our default numerical setup. 
However, a compromise solution between accuracy and computing sustainability could be identified for the
case ECP $\sqrt{N_{\rm pix}}= 131072$, where the reconstruction is well within percent level
with respect to the highest resolution case up to $\ell\lesssim 8000$.

As it can be expected from the results of the previous sections, similar conclusions can be drawn for the shear E and B-modes and ${\bf d}^{\rm E,eff}$, showing that the code is able to simulate the
expected signal with a high precision. We note however that the B-modes of the effective displacement did not reach a numerical stability with the same level of precision. For this field the numerical convergence between the cases adopting a resolution of $\sqrt{N_{\rm pix}}=131072$ and $\sqrt{N_{\rm pix}}=262144$ is of the order of 10\%. As confirmed by the consistency relation analysis of the previous section, this field seems to be the most sensitive to resolution effects. Even though this level of precision is likely enough for all practical applications, we conservatively decided not to use this quantity for the scientific analysis of the following sections and derive the effective displacement from the $\psi-\kappa$ and $\Omega-\omega$ consistency relations.
\subsection{Testing the N-Body simulation}\label{subsec:nbody_res}
\begin{figure}[!htb]
\centering
\includegraphics[width=.5\textwidth]{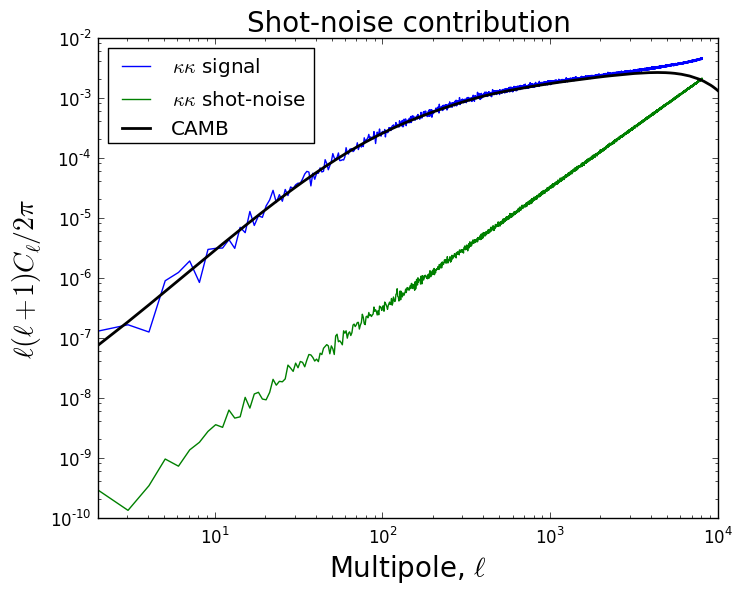}\includegraphics[width=.5\textwidth]{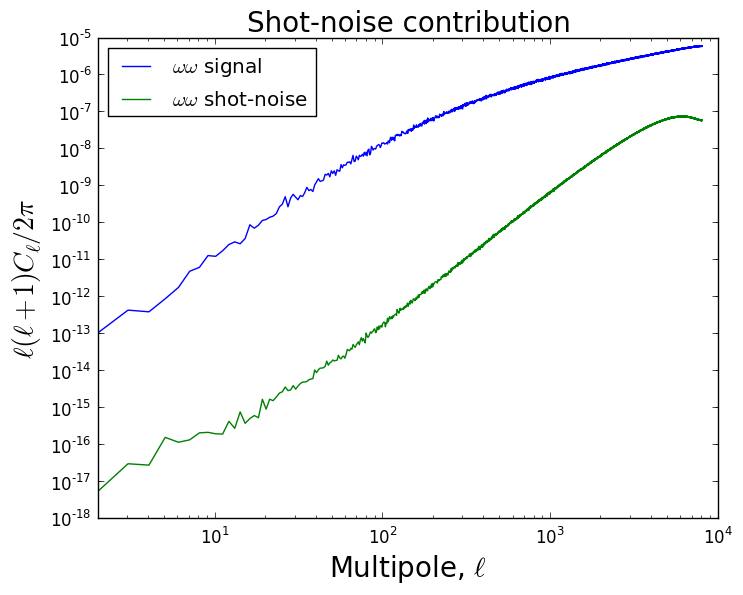}
\caption{\emph{Top}: Angular power spectrum for the scalar field
 convergence (\emph{left}) $\kappa$ and rotation (\emph{right}) $\omega$, both
 signal and shot-noise contribution. Black
 line is the \lstinline!CAMB! reference. 
}
\label{fig:shotnoise}
\end{figure}

\begin{figure}[!htbp]
\centering
\includegraphics[width=.49\textwidth]{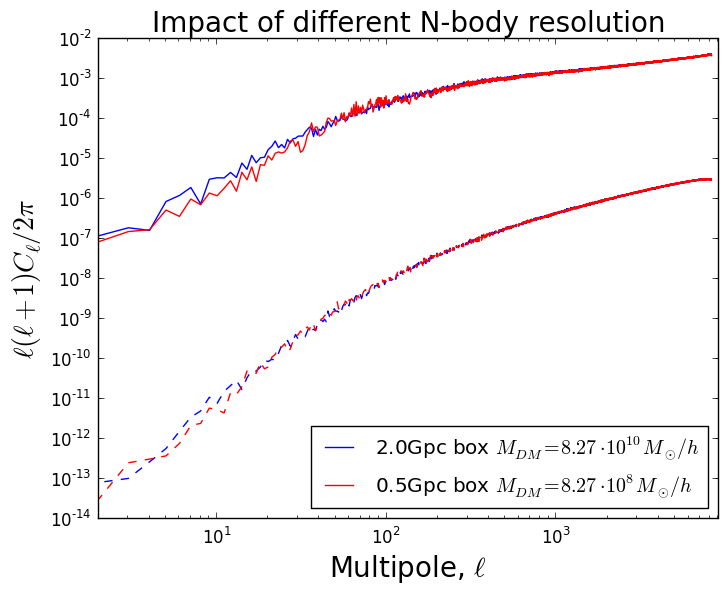}
\includegraphics[width=.49\textwidth]{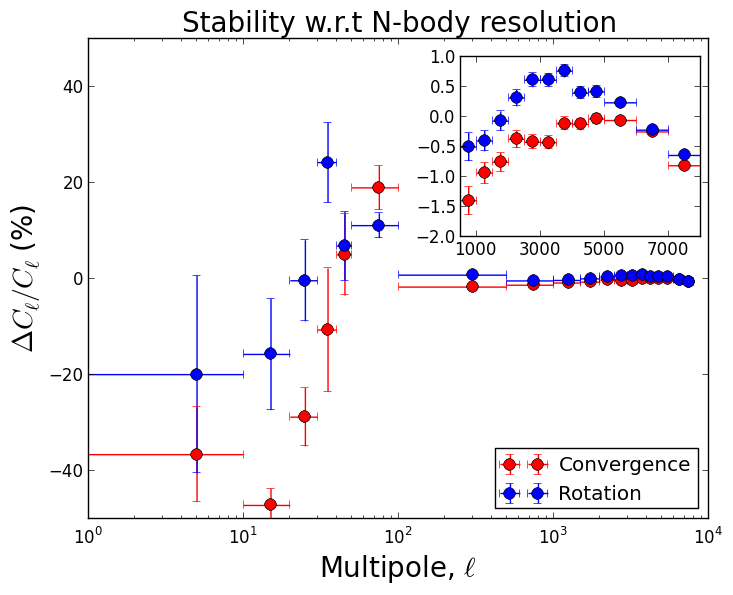}
\caption{\emph{Left}: angular power spectrum of lensing convergence (solid) and rotation (dashed) fields obtained from the DEMNUni simulation. Results obtained with a Millennium-like N-body simulation are displayed in red. \emph{Right}: binned fractional difference of power spectra obtained with the DEMNUni and Millennium-like simulations. Error bar represent the error on the mean of points following within the bin.}
\label{fig:nbodyres}
\end{figure}

Since beyond-Born effects are expected to show up as small angular scales where non-linear scales of the matter distribution contribute the most, it is important to quantify the impact of the resolution of the N-Body itself on the final result. 
In analogy to what we did in our previous work \cite{Calabrese14}, we first characterized the impact of the shot-noise generated from the discrete matter distribution inside the box after projection on the 2D sphere. We used for this purpose Monte Carlo Gaussian realization of the surface mass density field derived from the shot-noise power spectra in each shell \cite{Fosalba08}

\begin{equation}
C_{\ell}^{\kappa\kappa, \textrm{Shot-noise} (k)} = \frac{9H_{0}^{4}\Omega^{2}_{m,0}}{4c^{4}}\delta\chi \frac{1}{\bar{n}_{k}}\left(\frac{\chi_{s}-\chi_{k}}{\chi_{s}\chi_{k}}\right)^{2}
\end{equation}
\noindent
where $\bar{n}_{k}$ is the particle density in the k-th shell. 
We then used these maps as an input of the raytracing algorithm and propagated the shot-noise induced uncertainties on all the simulated lensing fields.
In Fig.~\ref{fig:shotnoise} we show the shot-noise contribution to
$\kappa$ and $\omega$. From this analysis it is evident that the excess of power at small angular scales that is observed in the recovered convergence power spectrum when compared with the CAMB results based on HALOFIT, is compatible with a noise bias induced by the simulation shot noise. We note however that
even if the shot-noise affects at 10\% level the convergence and shear E-modes, it is negligible for the rotation and
shear B-modes. This can be explained as the physical origin of those
signals is the coupling of subsequent deflections along the line of sight. A random realization of the shot noise has in fact a very small probability to
replicate such lens-lens configuration, which is indeed a peculiar and
interesting feature of the matter distribution from which the rotation
field arises.
\noindent
In addition, we also investigated whether the
simulation resolution for high value of $k$  significantly affected our results. In particular we
used as a benchmark case a Millennium-like simulation (500 Mpc boxsize, $2048^{3}$ dark matter particles) that adopts the same background
cosmology and amplitude of matter perturbations of the baseline N-Body simulation described so far but having different initial conditions. In Fig.~\ref{fig:nbodyres}, we show the convergence and rotation power spectrum extracted from
the Millennium-like and the baseline simulation. Even though it was not possible to run the Millennium-like simulation with the same initial conditions of our baseline setup, the agreement between the results recovered with these two simulations is remarkable. In particular we can see that dark matter halos of lower mass resolved in the Millennium-like simulation impact the convergence and rotation power spectrum below the 1\% level for $\ell>1000$. The only noticeable difference appears at large angular scales where the Millennium-like based results display a lack of power. This is expected since the box size is four
times smaller than the standard case, resulting in a more pronounced lack of power on scales larger than the boxsize, as noted in \cite{Calabrese14, Carbone09}

\subsection{Algorithm robustness}
\begin{figure}[!htb]
 \centering
\includegraphics[width=.5\textwidth]{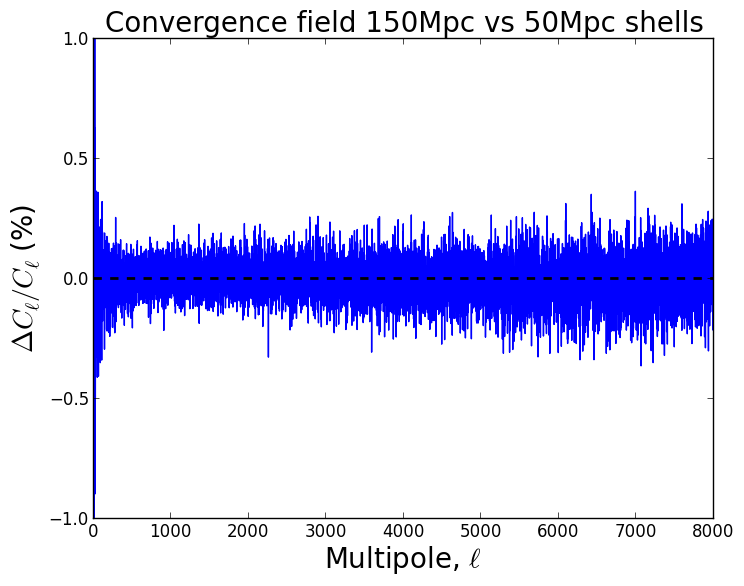}\includegraphics[width=.49\textwidth]{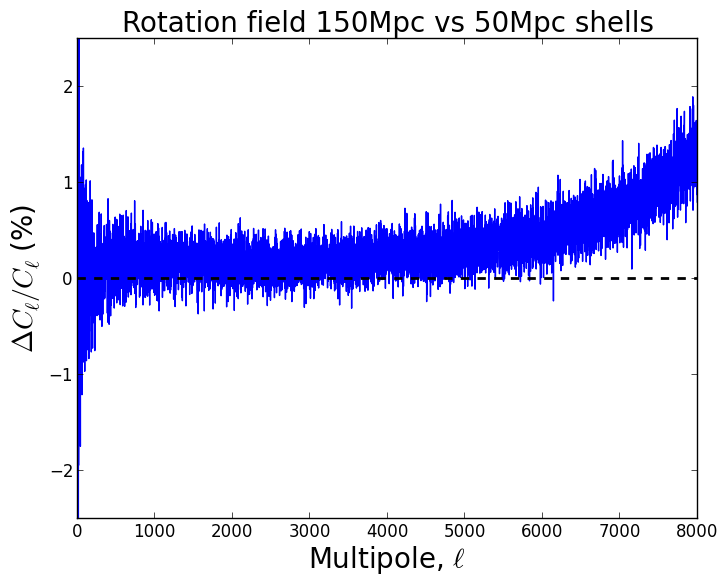}
 \caption{Fractional difference for the angular power spectra of the convergence (\emph{left}) and rotation (\emph{right}) obtained using 62 lens-planes (150 Mpc$/h$ width matter shells) and for an
  increased number of lens-planes (141) obtained slicing the lightcone in 50 Mpc$/h$ width
  matter shells.}
 \label{fig:number_lens_effec_e}
\end{figure}

A standard back-of-the-envelope calculation predicts that a CMB photon undergoes a total of about 50 independent deflections from large scale lenses (i.e. of $300h^{-1}$Mpc) during its journey from the last scattering surface towards us \cite{LewisChallinor06}. However, when considering lensing on smaller scales, there could be more deflections along the line of sight and thus it is important to test the stability of the simulation results against the number of employed lensing planes. We note that for an accurate modeling of post-Born correction, a relatively low number of planes could be sufficient. Post-Born effects require, in fact, a fairly large separation between the background and forward planes to have any impact on the lensing signal. There are no post-Born effects for lenses right next to each other, i.e the $f_K(\chi_{k+1} - \chi_k)$ factor of~\eqref{eq:fulllenseq_dis} approaches zero. \\*
In the previous sections we presented the results of raytracing simulations employing 62 lensing planes as a reference setup. This number corresponds to the
number of N-body snapshot at our disposal and thus limits the number of randomization the we have to perform in order to fill the full past lightcone. To assess the robustness of the results with respect to this specific choice in the number of lensing planes, we constructed the same lightcone realization employed throughout this work but then adopted a thinner matter shells slicing. In the benchmark
case, the 62 outputs of the simulation have been translated into 62
matter shells of (average) comoving width of about 150 Mpc$/h$. Instead, for
this test, we sliced the lightcone into 141 matter shells of 
(average) comoving width of about 50 Mpc$/h$. This scenario
traces exactly the same matter distribution and evolution as the benchmark
case, but it models
the lensing signal with an increased number of deflections. In
Fig.~\ref{fig:number_lens_effec_e} we compared the results obtained with these two approaches. We found that the convergence signal
is stable at better than 0.5\% level, however, the rotation signal changes by a factor
of 1.5 \% on scales close to $\ell=8000$. The latter is expected, because if we increase the
number of deflections, we expect the signal that depends on the
lens-lens configuration to increase as well. However such difference is barely above the amplitude of the cosmic variance at those scales and the number of lens planes increased at the same time by more than a factor of 2. The number of lens planes employed in our baseline simulation setup seems therefore sufficient to resolve beyond-Born corrections, induced by the lens-lens coupling, with reasonable accuracy. 

\section{Comparison with theoretical predictions}\label{sec:analytical-results-comparison}
After having confirmed the robustness of our result
from a numerical point of view, in this section we compared the recovered signal with theoretical
predictions based on the perturbative approach to the lens equation and beyond-Born correction \cite{HirataSeljak2003, CoorayHu02, KrauseHirata10, pratten2016, lewis-pratten2016, marozzi2016, marozzi2016dec}. 
\subsection{Convergence corrections}
\begin{figure}[!htbp]
\centering
\includegraphics[width=.495\textwidth]{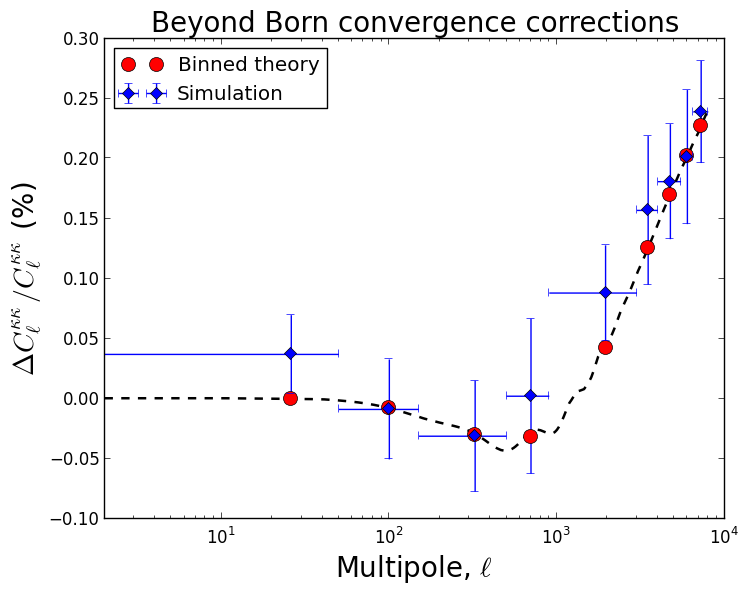}
\includegraphics[width=.49\textwidth]{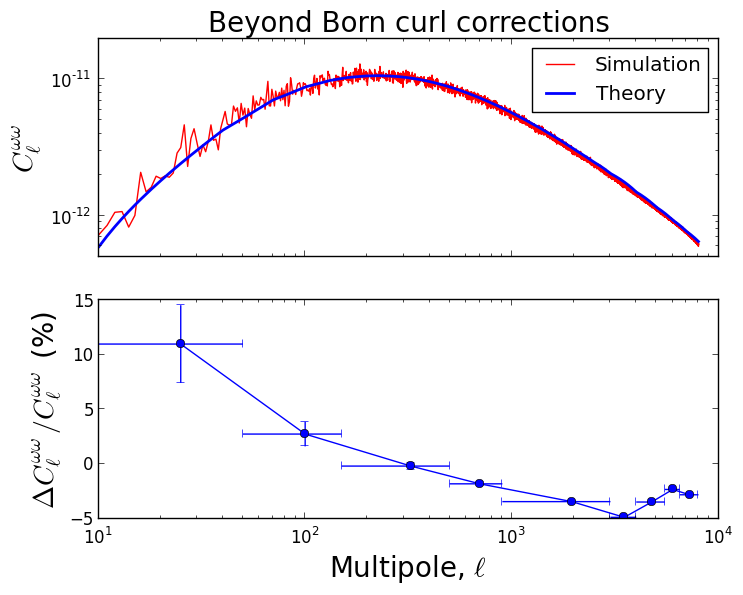}
\caption{\emph{Left}: fractional difference of first-order and beyond-Born lensing convergence power spectra. Black dashed line shows the analytical expectation for Planck 2013 cosmology. Red bullet points show the binned version of this curve. Simulation results are displayed in blue. Bullet points show the mean value inside each bin while error bars accounts from the error on the mean itself. \emph{Right}: analytical and simulation estimate of the rotation power spectrum (top) and their fractional difference (bottom).}
\label{fig:sims_vs_antony}
\end{figure}

\begin{figure}[!htbp]
\centering
\includegraphics[width=.45\textwidth]{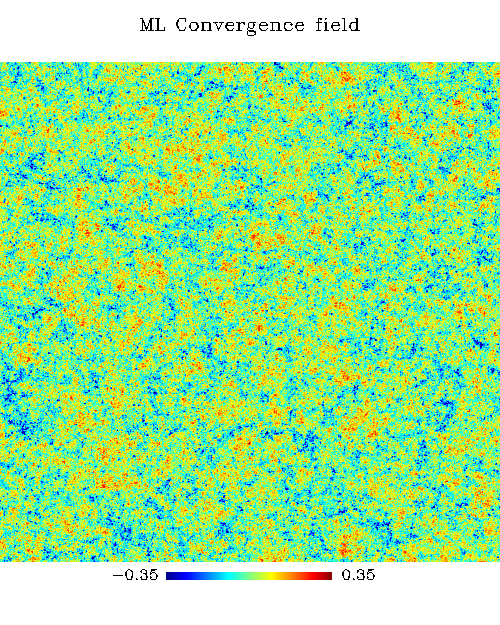} \quad\includegraphics[width=.45\textwidth]{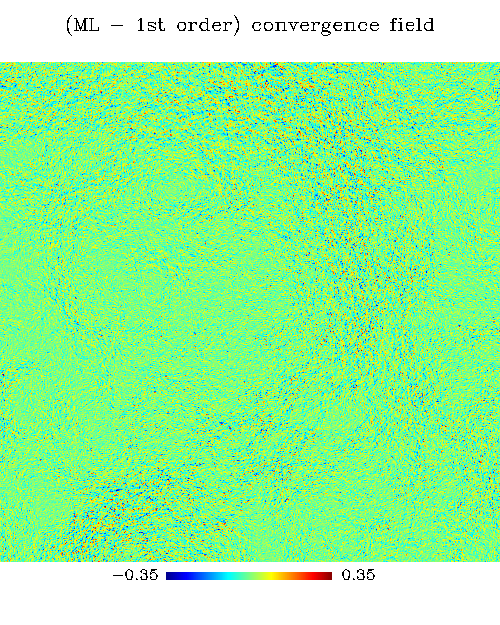} \\
\includegraphics[width=.49\textwidth]{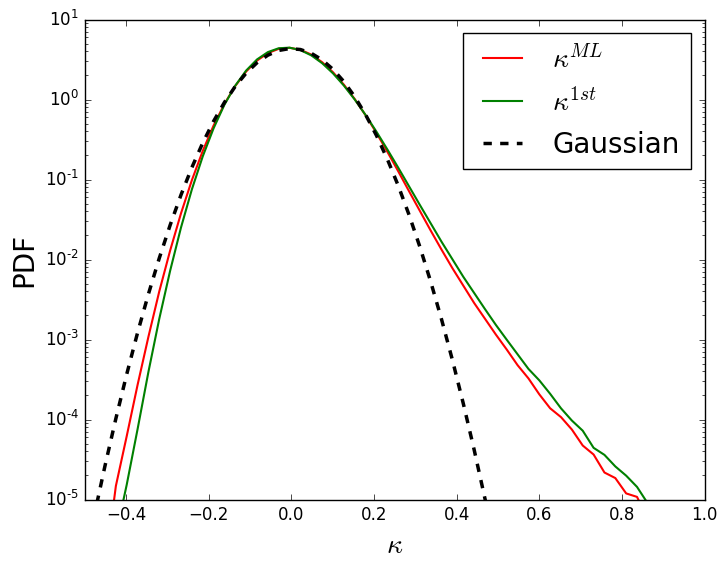}
\includegraphics[width=.49\textwidth]{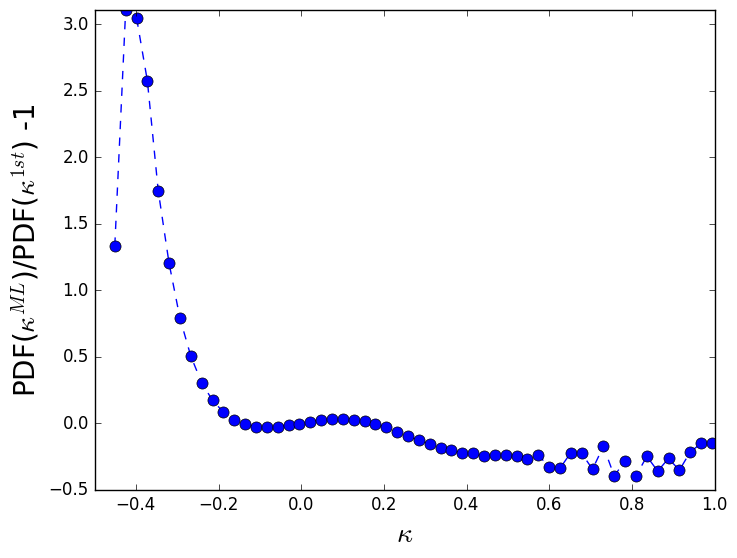} 
\caption{\emph{Top}: map of the $\kappa^{\rm ML}$ field (\emph{left}) and of $\kappa^{\rm ML}-\kappa^{\rm 1st}$ (\emph{right}). \emph{Bottom}: PDF of the maps of the top panel (left). The dashed black line shows the Gaussian approximation to the PDF defined by the variance of $\kappa^{\rm 1st}$. The fractional difference between the PDF of $\kappa^{ML}$ and $\kappa^{\rm 1st}$ is shown in the right panel. Beyond-Born corrections tend to reduce the higher order moments of the PDF.}
\label{fig:kappa-pdf}
\end{figure}

\begin{figure}[!htbp]
\includegraphics[width=.5\textwidth]{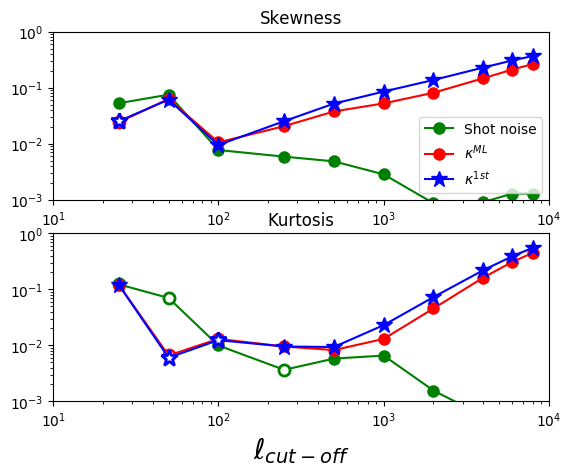}\includegraphics[width=.5\textwidth]{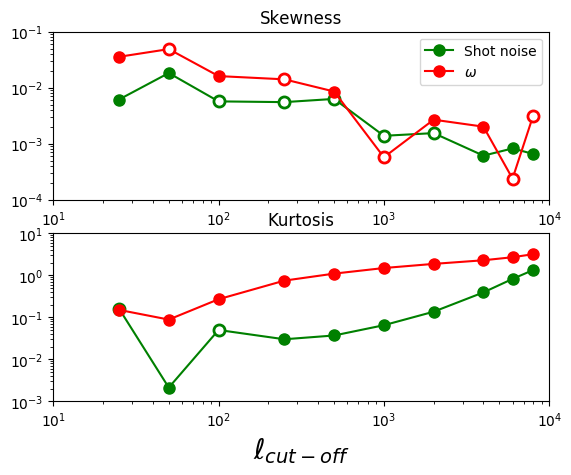}
\caption{Higher order moments of 1-point PDF of $\kappa$ (\emph{left}) and $\omega$ (\emph{right}) maps including power only on angular scales $\ell<\ell_{\rm cut-off}$. Negative values are shown with empty markers. Typical values for a shot-noise realization in these fields are shown in green.}
\label{fig:non-gauss-vs-lmax}
\end{figure}

As noted in \cite{Calabrese14} and confirmed analytically by \cite{pratten2016} (hereafter PL16), the impact of beyond-Born corrections on the CMB lensing convergence signal is small and the first-order result dominates. A direct and accurate comparison between analytical predictions and numerical results is not straightforward because differences might be easily hidden in the cosmic variance scatter. The latter cannot be reduced given the impossibility of performing a full Monte Carlo analysis with N-body simulation of the size employed in this work. 
In Fig.~\ref{fig:sims_vs_antony} we compared the fractional difference of the lensing convergence power spectrum beyond-Born and the first-order result with the theoretical prediction computed with the numerical implementation of the perturbative corrections presented in PL16\footnote{\url{http://cmbant.github.io/notebooks}}. Despite the fact that it was not possible to investigate quantitatively the agreement $\ell$ by $\ell$ for the reasons outlined above, the binned version of these curves shows a good agreement between simulations and analytical predictions, with a clear trend of an increase of power at small angular scales and a deficit of power at multipoles of few hundreds. \\*
\noindent
For sake of completeness we investigated the impact of beyond-Born corrections also in the real domain. In particular, we looked at the moments of the convergence maps obtained in the first-order approximation $\kappa^{\rm 1st}$ and in the beyond-Born regime $\kappa^{\rm ML}$.
The quantity $\Delta\sigma^2_{\kappa}= \sigma^{2}_{\kappa^{\rm ML}} - \sigma^{2}_{\kappa^{\rm 1st}}$ can in fact be directly connected to the beyond-Born correction on the convergence power spectrum $\Delta C^{\kappa\kappa}_{\ell}$ using the well-known relation between the variance of a scalar field and its power spectrum 
\be
\Delta\sigma_{\kappa}^{2, \rm Theory} \equiv \left(\sigma^2_{\kappa^{\rm ML}} - \sigma^2_{\kappa^{\rm 1st}}\right)^{\rm Theory} =
\sum^{\ell_{\rm max} }_{\ell=0} \frac{(2\ell
 +1)}{4\pi}\Delta C^{\kappa\kappa}_{\ell}.
\label{eq:delta-sigma-theory}
\ee
We estimated the same quantity from the simulated $\kappa^{\rm 1st}$, $\kappa^{\rm ML}$ maps, $\Delta\sigma_{\kappa}^{2,\rm Sim}$, and compared it to the predictions of~\eqref{eq:delta-sigma-theory} derived using the analytical expectations for $\Delta C^{\kappa\kappa}_{\ell}$. To minimize the impact of numerical effects towards the end of the band-limit of the signal, we filtered the simulated maps setting to zero all the $\ell> 8000$ modes and fixed the upper limit of the sum in Eq.~\eqref{eq:delta-sigma-theory} accordingly. We found that $\Delta\sigma_{\kappa}^{2, \rm Sim} = 7.25 \cdot 10^{-6}$ and $ \Delta\sigma_{\kappa}^{2, \rm Theory} = 7.00\cdot 10^{-6}$
respectively, thus the analytical predictions of beyond-Born correction match to a precision of roughly $3\%$ the results derived in the simulations for this specific realization of the matter distribution.\\*
The recent work of PL16 noted that beyond-Born corrections generate a non-negligible amount of non-Gaussianities, in particular in the $\kappa\kappa\kappa$ bispectrum, and that those might be detected in the near future unlike corrections on $C_{\ell}^{\kappa\kappa}$. For this reason, we measured the variation of skewness and kurtosis of the 1-point PDF of $\kappa^{\rm 1st}$ and $\kappa^{\rm ML}$. The 1-point PDF of $\kappa^{\rm ML}$ and $\kappa^{\rm 1st}$ are shown in Fig.~\ref{fig:kappa-pdf} together with the Gaussian approximation of $\kappa^{\rm 1st}$ defined as a Gaussian PDF having mean and variance equal to the one measured on the simulated map itself. Both $\kappa^{\rm ML}$ and $\kappa^{\rm 1st}$ PDF follow a Das and Ostriker model as it could have been expected from the analysis of Sec.~\ref{sec:maps-systematics}. 
As discussed above, the variances of the two distributions are very similar. However their skewness and kurtosis change by a factor of roughly 30\% when beyond-Born corrections are considered (see table~\ref{table:kappa-moments} for a summary of these values). In particular, as shown in Fig.~\ref{fig:kappa-pdf}, the beyond-Born correction tends to modify mainly the high and low ends of the distribution. In Fig.~\ref{fig:non-gauss-vs-lmax} we show the measurements of the higher order moments of $\kappa^{\rm ML}$ and $\kappa^{\rm 1st}$ and $\kappa^{\rm shot-noise}$ obtained filtering the maps in order to retain only the modes $\ell \leq \ell_{\rm cut-off}$. We performed this test to assess whether large-scales artifacts induced by the N-body boxsize replication (especially in the high redshift shells) or shot noise effects dominated the values of the skewness and kurtosis of the map. As it is visible from the figure, the higher order moments are dominated by non-Gaussianities located at progressively smaller angular scale, consistent with non-linearities arising in the gravitational evolution of the matter distribution. In particular comparing the values of skewness and kurtosis of the first-order and ML maps we can clearly see the onset of the change in the amount of non-Gaussianities due to beyond-Born corrections starting at scales $\ell\approx 250$ and a negligible shot-noise contribution to these values for scales $\ell>1000$. Although a quantitative assessment of the precision of the analytical predictions for the amount of non-Gaussianities produced by beyond-Born corrections is beyond the scope of this paper, we qualitatively confirm the findings of PL16: beyond-Born corrections affect more the non-Gaussian part of $\kappa$ than its power spectrum. PL16 also noted that the specific signature of beyond-Born corrections to the $\kappa\kappa\kappa$ bispectrum depends on the specific triangle configuration and redshift. The effect consists mainly in a combination of a reduced lensing efficiency generated by lens coupling (which suppresses the bispectrum of the matter distribution) and additional distortions due to deflections across the direction where the gravitational potential changes. These two contributions have an opposite sign and their relative amplitude changes with redshift.  The results of our simulations suggest that the leading effect of beyond-Born corrections consists in an overall suppression of the amount of non-Gaussianities in the CMB lensing convergence. This result seems to agree with the naive expectations from the results of PL16 (see, e.g., their Fig. 4 where the beyond-Born $\kappa\kappa\kappa$ bispectrum is mostly negative). However, being the skewness the sum of all the configurations of the $\kappa\kappa\kappa$ bispectrum, we cannot make a quantitative statement on the validity of the analytical predictions of this signal. Thus, we postpone a more quantitative analysis of this aspect on our simulated map to a future work. 

\begin{table}[!htbp]
\centering
\begin{tabular}{c|ccc}
\hline\hline
Map& Variance ($10^{-3}$)& Skewness & Kurtosis\\
\hline
$\kappa^{1\rm st}$ & 8.465 & 0.370 & 0.554 \\
$\kappa^{\rm ML}$ & 8.472 & 0.266 & 0.450\\
$\omega$ & $6.87\cdot 10^{-3}$ & $3.184\cdot 10^{-3}$ & 3.185\\
$\kappa^{\rm ML}$ shot-noise & 1.070 & $1.289\cdot 10^{-3}$ & $-3.865\cdot 10^{-4}$\\
$\omega$ shot-noise & $6.357\cdot 10^{-5}$ & $6.570\cdot 10^{-3}$ & $1.337$\\
\hline\hline
\end{tabular}
\caption{Moments of the first-order and beyond-Born convergence and rotation maps (together with moments of the corresponding shot-noise maps) obtained filtering modes $\ell>8000$. }
\label{table:kappa-moments}
\end{table} 


\subsection{Lensing rotation corrections}
\begin{figure}[!htbp]
\centering
\begin{minipage}[c]{0.45\textwidth}
\includegraphics[width=\textwidth]{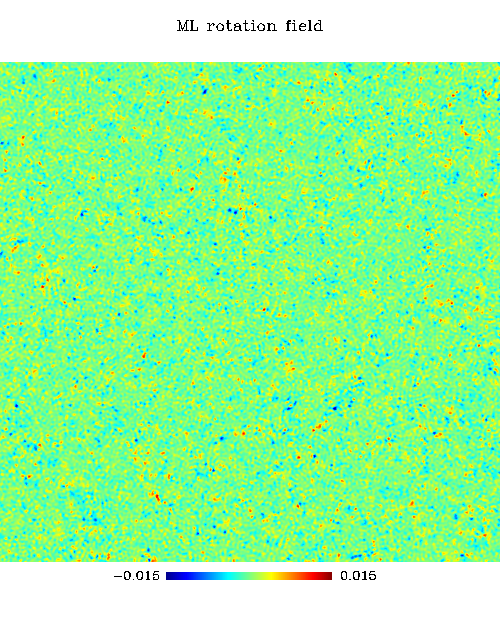}
\end{minipage}
\begin{minipage}[c]{0.54\textwidth}
\includegraphics[width=\textwidth]{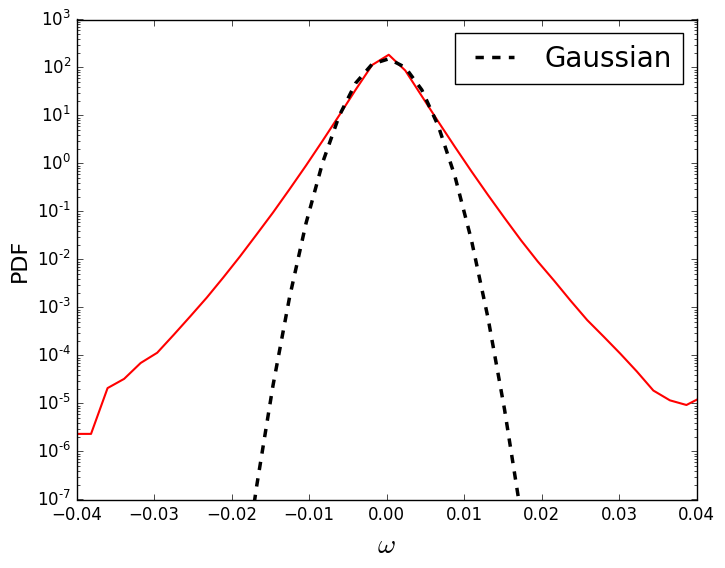}
\end{minipage}
\caption{Map of the $\omega$ field (\emph{left}) and its PDF (\emph{right}). The dashed black line shows the Gaussian approximation of the PDF defined by the variance of the $\omega$  map itself.}
\label{fig:rotation-pdf}
\end{figure}

One of the most characteristic signs of beyond-Born corrections is the introduction of an image rotation field. In Fig.~\ref{fig:sims_vs_antony} we show the comparison between the analytical expectation for the rotation power spectrum and the result extracted from the simulation. We note that, despite the procedure devised to define the band-limits for the Poisson equation resolution (see Sec.~\ref{subsec:magnif_spectra}) was gauged on the estimation the $\kappa$ field, $\omega$ is recovered with a similar level of accuracy. The results of the simulation agree to better than 5\% at all scales and to better than 1\% on scales around the rotation power spectrum peak. The largest discrepancy with analytical results can be observed at large angular scales ($\ell\lesssim 50$). Few factors may be the reason of this discrepancy on top of the uncertainties in modeling of the non-linear matter power spectrum. Analytical results have in fact been derived in flat-sky and using the Limber approximation, which becomes less accurate in these regimes. At the same time additional correlation may be introduced by boxsize replication during the map-making step of the algorithm (see Sec.~\ref{sec:algorithm:mapmaking}) and the excess might be related to this procedure.\\ 
Similarly to what we did in the previous section, we investigated whether the variance measured in the map matched the one computed from the theoretical $C_{\ell}^{\omega\omega}$. We found that this number agrees within 10\%, with a contribution of the shot noise of 1\% or less. On the contrary, if we look at the 1-point PDF of the $\omega$ map in Fig.~\ref{fig:rotation-pdf} we can see that the field is significantly non-Gaussian and the observed distribution is peaked and with long tails (see table~\ref{table:kappa-moments} for the moments of the distribution). Unlike the convergence case, where higher order moments never exceed a value of 1, the level of non-Gaussianities is significant. This is an additional source of possible discrepancies with the analytical calculations that so far have not addressed the impact of non-Gaussianities in the $\omega$ field alone. The skewness of the $\omega$ map is much smaller than its kurtosis. For this reason we suppose that the major contribution to the rotation field non-Gaussianities may come from the trispectrum. Since a large and positive trispectrum is usually sign of highly non-linear and local effects \cite{lewis-nongauss}, this signature might be related to the presence of massive and clustered halos in the maps that cause deviations from the purely weak lensing regime. This hypothesis should however be investigated quantitatively with dedicated tests and we leave this analysis to future work.\\* 
We finally note that the typical level of kurtosis expected for a shot-noise realization following the model used in Sec.~\ref{subsec:nbody_res} can account for 30-40\% of the amplitude of the kurtosis (see table~\ref{table:kappa-moments}). However, a simple analysis of the scale dependency of the kurtosis analogous to the one of the previous section showed that the shot-noise contribution is negligible (few percent) up to scales $\ell\approx 2000$ where the kurtosis is already significant and equal to 2 (see Fig.~\ref{fig:non-gauss-vs-lmax}). \\*
\noindent
Despite the prospect of measuring $\omega$ from the reconstruction of curl-deflections of CMB lensing are not promising \cite{pratten2016, sheere2016} - and thus analysis of its non-Gaussianities might not seem useful for upcoming experiments - we note that a curl-like deflection is very efficient in generating B-modes \cite{Padmanabhan2013} and might leave a detectable imprint on top of the standard lensing B-modes signal generated by the gradient mode. In particular, in the context of CMB lensing, a significant level of non-Gaussianities in the curl mode may affect the estimators used to reconstruct the CMB lensing and the curl deflection potential from CMB maps \cite{HuOkamoto02, Namikawa2013}. These in fact look for non-Gaussian signatures in the CMB maps left by lensing assuming the potentials are Gaussian. A first analysis of this aspect for the CMB lensing potential reconstruction has been reported in \cite{bohm2016}. A more detailed analysis might be important to understand the ultimate precision limit with which the lensing-induced contribution to the CMB B-modes signal can be subtracted from the measured signal to enhance the detection of their primordial counterpart. This is in fact ultimately related to the presence of curl deflections and to their peculiar statistical properties. \cite{HirataSeljak2003, Smith2012}. We will discuss these issues in more details in the following section. \\*
We finally note that despite $\kappa$ and $\omega$ are uncorrelated at first-order and thus $C_{\ell}^{\kappa\omega} = 0$, these fields are not independent. They in fact have a non-null $\kappa\kappa\omega$ mixed bispectrum, as explained in PL16. We discuss details of this correlation in Sec.~\ref{sec:cmb-lensing}.

\section{Lensed CMB beyond-Born approximation}\label{sec:cmb-lensing}
\subsection{Methodology}
\begin{figure}[!htbp]
\centering
\includegraphics[width=.45\textwidth]{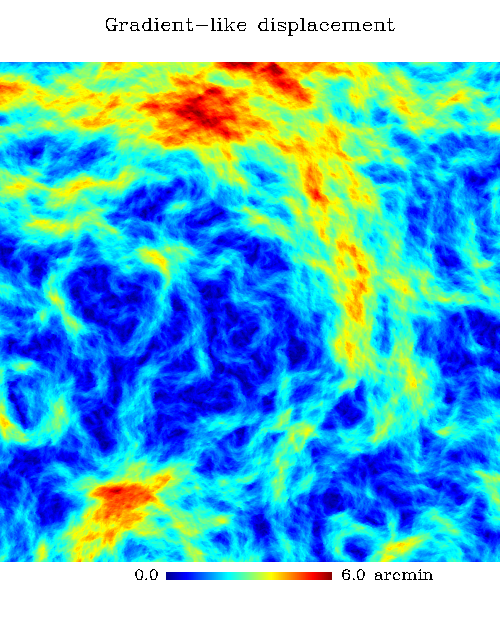}\qquad\includegraphics[width=.45\textwidth]{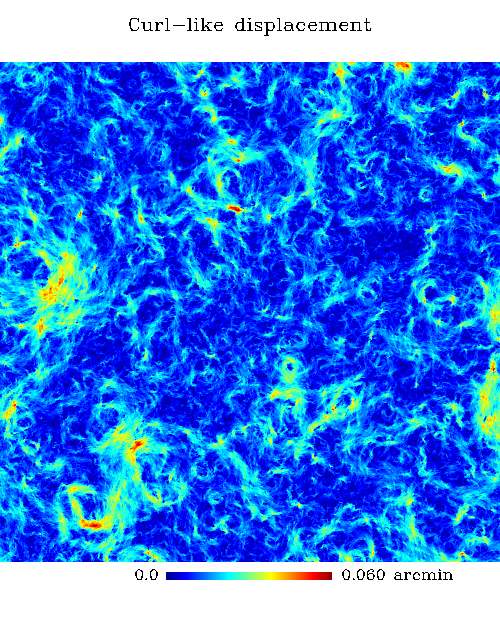} 
\caption{Magnitude of the deflection field generated from the gradient-like (\emph{left}) and curl-like (\emph{right}) deflections evaluated from the effective lensing and curl potentials of the DEMNUni simulation.}
\label{fig:grad-curl-displacement}
\end{figure}

In the previous sections we described the properties of the simulated CMB lensing convergence and rotation fields including beyond-Born corrections. In this section we use these quantities to evaluate the impact of beyond-Born corrections on the CMB observables commonly used to estimate cosmological parameters, i.e. the angular power spectra of its temperature and polarization anisotropies.\\* 
For this purpose, it is convenient to adopt an effective approach where CMB photons undergo a single deflection as done in the standard Born approach but using a deflection field $\mathbf{d}^{\rm eff}$ synthesized from the $\kappa^{\rm ML}$ and $\omega$. The spin-1 E and B harmonic coefficients of this field are
\begin{equation}
_{1}d^{\rm E, eff}_{\ell m} = -2\frac{\kappa^{\rm ML}_{\ell m}}{\sqrt{\ell(\ell+1)}} \qquad _{1}d^{\rm B, eff}_{\ell m} = -2\frac{\omega_{\ell m}}{\sqrt{\ell(\ell+1)}}.
\label{eq:effective-displacement}
\end{equation}
Because the consistency relations between $\kappa^{\rm ML}$  and $\psi$, as well as $\omega$  and $\Omega$, are satisfied with good accuracy (see Sec.~\ref{sec:consistency}), this is equivalent to using the ${\bf d}^{\rm eff}$ field but minimizing the impact of residual aliasing. Moreover this procedure is analogous to propagating the CMB sky through the shells as done in \cite{Calabrese14}. However, it allows to minimize the numerical effects induced by the pixel remapping operation that otherwise would have to be performed each time the ray trajectory is deflected by a lens plane in the multiple lens approach (see discussion in \cite{Calabrese14, Fabbian13}). A snapshot of the $\mathbf{d}^{\rm eff}$ field is shown in Fig.~\ref{fig:grad-curl-displacement}.
\noindent
For sake of completeness, we note that when we lens CMB polarization we have to take into account the parallel transport of the basis of the tangent plane between the unlensed and lensed direction on the 2-sphere each time a light ray is deflected. This is a direct consequence of the fact that CMB polarization is a spin-2 field defined on the sphere. The parallel transport operation can be recast as a rotation of the components of the polarization field after each deflection by a proper angle $\delta$, which is used to rotate the polarization components back into the initial reference frame. This angle can be expressed in terms of the deflection field components and the incoming direction of the light ray \cite{Lewis05, challinor-chon}. Every time we computed a ray deflection at each lensing plane in the raytracing, we evaluated at the same time the corresponding rotation angle $\delta^{(i)}$, and then updated subsequently a global effective rotation angle $\delta^{\rm raytracing}$ that describes the total rotation of the polarization components after the i-th deflection. To compute $\delta^{(i)}$, we adopted the formulae presented in the appendix of \cite{Lewis05} and used the deflection field ${\bf d}^{(i)}$ defined in Sec.~\ref{sect:raytracing-remapping}. This takes into account the incidence angle and the actual deflection at the plane crossing and it is equal to the effective displacement of the angular position of the ray seen by an observer.
In the following we will compare $\delta^{\rm raytracing}$ with the angle derived solely from the parallel transport of the ray from its unlensed to its final direction. We compute the latter using the effective displacement described above, and denote this quantity as $\delta^{\rm transport}$ in the following. We will comment more on this point in the following sections. \\* 
Finally, we extracted the deflection field $\mathbf{d}^{\rm eff-1st}$ using the effective lensing potential recovered from the first-order convergence. This deflection field is used to derive the lensed CMB power spectra that serve as a benchmark for the results based on the Born approximation.

\subsection{Impact on lensed CMB power spectra}\label{results:power-spectra}
\begin{figure}[!htbp]
\centering
\includegraphics[width=\textwidth]{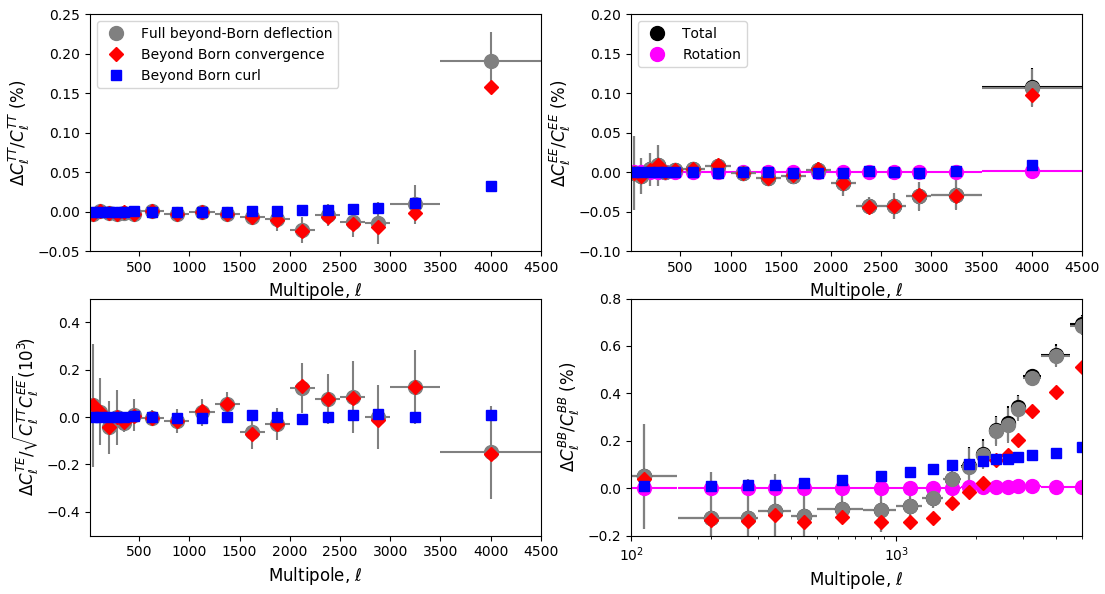} 
\caption{Impact of beyond-Born corrections on the lensed CMB power spectra with respect to the first-order results (gray). Error bars show the standard deviation of 100 Montecarlo realization of lensed CMB sky using the DEMNUni lightcone and different unlensed CMB sky. Single contributions to the total correction on $C_{\ell}$ coming from the different components of the beyond-Born correction are highlighted in different colours. Polarization rotation corrections are derived using the $\beta^{\rm rotation}$ angle measured from the simulation (see Sec.~\ref{results:power-spectra}).}
\label{fig:delta-cl-cmb-beyond-born}
\end{figure}

\begin{figure}[!htb]
\centering
\includegraphics[width=.5\textwidth]{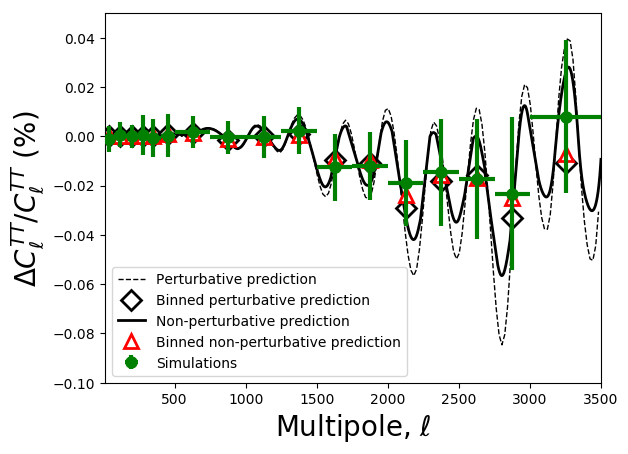}\includegraphics[width=.5\textwidth]{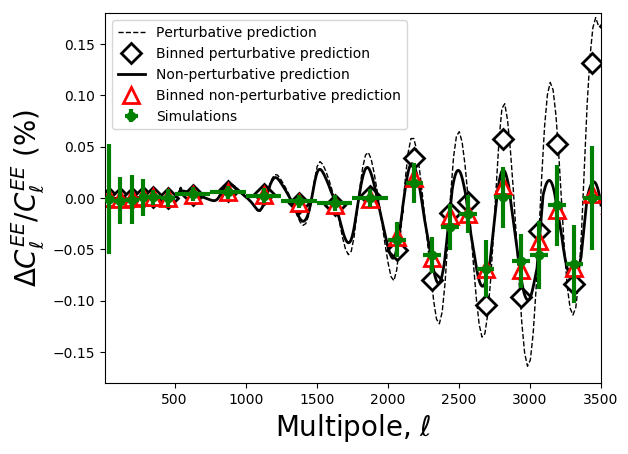}
\caption{Comparison of the analytical predictions of beyond-Born corrections of \cite{marozzi2016dec} (black dashed) with numerical simulations of this work (solid dots) for the CMB temperature (left), E-modes (right) power spectra.  Error bars correspond to the MC mean dispersion. The results of the non-perturbative calculations of \cite{lewis-pratten2016} are shown as black solid lines. The agreement between simulations and theoretical predictions is very good and improves when including the non-perturbative results for the E-modes power spectra at small scales.}
\label{fig:t-e-modes-theory}
\end{figure}

\begin{figure}[!htb]
\includegraphics[width=.5\textwidth]{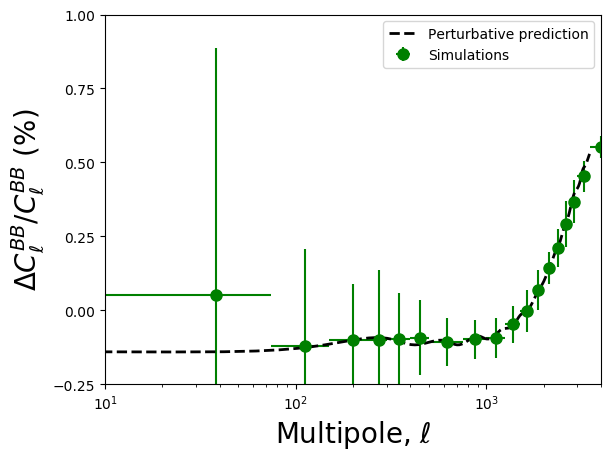}\includegraphics[width=.5\textwidth]{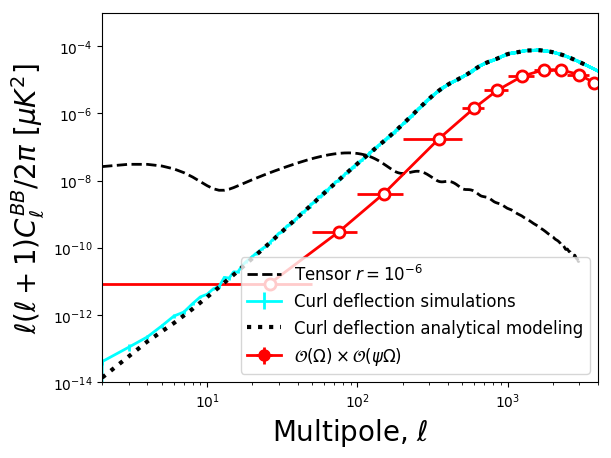}\\
\includegraphics[width=.5\textwidth]{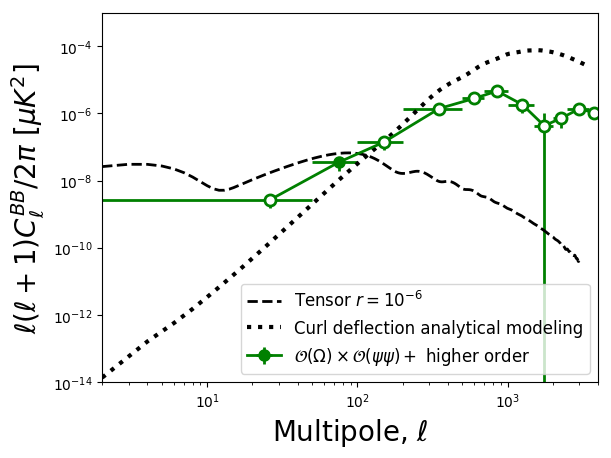}\includegraphics[width=.5\textwidth]{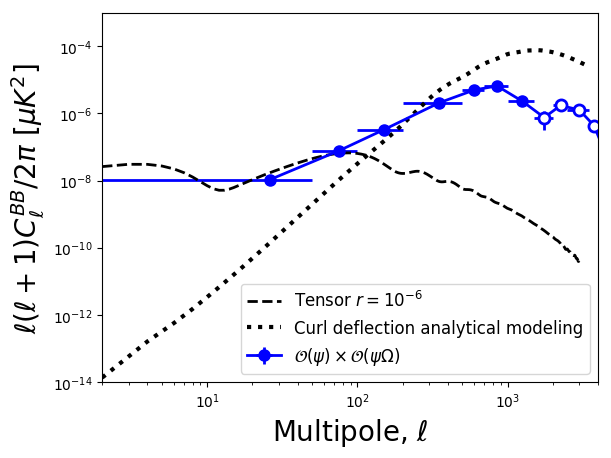}
\caption{Comparison of the analytical predictions of beyond-Born corrections on the B-modes power spectrum of \cite{marozzi2016dec} (dashed black) with numerical simulations of this work (solid dots). Error bars correspond to the MC mean dispersion. The top right and bottom panels show the absolute value of the contributions to the lensed B-modes power spectrum extracted from the simulations, and negative points are denoted with an empty dot. In the top right panel we show the contributions induced by the curl deflections alone (cyan) and by the previously neglected $\Omega\Omega\psi$ bispectrum contribution. Additional higher-order contributions to the lensed B-modes coming from contractions of the mixed bispectrum $\psi\psi\Omega$ are shown in the bottom panels. The so-called correlation term (Eq.~\eqref{eq:bmodes-correlation}) is shown on the bottom left panel while the so-called mixed term (Eq.~\eqref{eq:bmodes-mixed}) is shown in the bottom right panel. The theoretical predictions of the B-modes signal induced by Gaussian curl deflections alone derived using the PL16 formalism and the $C_{\ell}^{\Omega\Omega}$ extracted from the simulations is shown as a black dotted line. B-modes generated by tensor perturbations having $r=10^{-6}$ are shown as dashed black lines for reference in the top right and bottom panels.}
\label{fig:zoom-bmodes}
\end{figure}

Once we extracted the effective displacement field, we lensed the incoming CMB photons using the pixel based method implemented in the \lenstwo\ code \cite{Fabbian13} with an oversampling factor of 8. For these simulations we filtered all the power in the displacement field for scales $\ell>8000$ and set the band-limit for the unlensed CMB to $\ell^{\rm CMB}_{\rm max}={8000}$. This setup is sufficient to obtain 0.1\% accuracy on the lensed temperature, E and B-modes power spectra up to $\ell\approx 6000$ \cite{Fabbian13}.\\* 
In order to measure the impact of the beyond-Born corrections, we produced a set of simulated CMB maps where we lensed 100 independent unlensed Gaussian CMB realizations using four different deflection fields: $\mathbf{d}^{\rm eff}$, $\mathbf{d}^{\rm eff-grad}$, $\mathbf{d}^{\rm eff-curl}$ and $\mathbf{d}^{\rm eff-1st}$. $\mathbf{d}^{\rm eff-grad}$ and $\mathbf{d}^{\rm eff-curl}$, in particular, were computed setting either B or E modes to zero respectively in Eq.~\eqref{eq:effective-displacement}.  
The CMB maps lensed with $\mathbf{d}^{\rm eff-1st}$ are instead used as reference quantities for results based on the Born approximation. We then compared each of the lensed CMB power spectra computed with the four different displacements in order to isolate the impact of specific beyond-Born corrections.\\* 
Following \cite{pratten2016, lewis-pratten2016, marozzi2016dec} we can classify the corrections to the lensed CMB power spectra into two types. The first involves corrections to the deflection field, which can be further divided into the contributions coming from the beyond-Born curl and gradient components of the deflection, or contributions from the higher order correlations of the two effective lensing and curl potentials. Since the consistency relations are valid, and following the terminology of PL16, we refer to these higher order effects as $\kappa\kappa\kappa$ and $\kappa\kappa\omega$ (or mixed) bispectra. In our simulations, lensed CMB maps obtained with $\mathbf{d}^{\rm eff}$ hold all beyond-Born effects coming from the combined action of both gradient and curl deflections as well as their higher order correlations $\kappa\kappa\kappa$ and $\kappa\kappa\omega$. $\mathbf{d}^{\rm eff-grad}$ and $\mathbf{d}^{\rm eff-curl}$ contain only beyond-Born corrections on the gradient deflections (including the effect of  $\kappa\kappa\kappa$ bispectrum) and curl deflections respectively. \\* 
The second type of beyond-Born corrections affects only the polarization field and it consists in the rotation of the polarization tensor about its propagation direction by an angle $\beta^{\rm rotation}$. This effect appears when considering gravitational perturbations induced by higher-order scalar perturbations or vector and tensor perturbations at all perturbative orders \cite{dai2014}. 
The total beyond-Born lensed polarization field $\tilde{P}\equiv\tilde{Q}+i\tilde{U}$, where Q and U are the CMB Stokes parameters, is then related to the unlensed field $P$ as
\begin{equation}
\tilde{P}(\boldsymbol\theta) = e^{-2i\beta^{\rm rotation}}\left[e^{-2i\delta^{\rm transport}}P(\boldsymbol\theta - \mathbf{d}^{\rm eff})\right].
\label{eq:birefringence}
\end{equation}
\noindent
In the following, we denote with the superscript \emph{rotation} the quantities including this additional rotation of the lensed polarization field. We measured $\beta^{\rm rotation}$ from our simulation as the additional rotation of the coordinate basis on the sphere observed when performing raytracing with respect to the rotation angle corresponding to the simple parallel transport of the ray. Following the notation of the previous section, this reads $\beta^{\rm rotation} = \delta^{\rm raytracing} - \delta^{\rm transport}$. \\*
Because each CMB realization differs only by the lensing deflection used to lens the CMB, we can isolate the effect of each type of correction computing simply fractional differences.  For the sake of brevity, we defined $\Delta^{x, y}_{z}(C_{\ell}^{XY})$ as the difference between the lensed CMB power spectrum $C_{\ell}^{XY}$ obtained using the $x$ and $y$ deflection fields described above, divided by the same CMB power spectrum obtained with the $z$ deflection field. In particular, to display all the beyond-Born corrections we focused on the following quantities:
\begin{enumerate}
\item $\Delta^{\rm eff,eff-1st}_{\rm eff-1st}(C_{\ell}^{XY})$ shows the amplitude of all beyond-Born corrections on the deflection field. These are shown as gray dots in Fig.~\ref{fig:delta-cl-cmb-beyond-born}.
\item $\Delta^{\rm eff-grad, eff-1st}_{\rm eff-1st}(C_{\ell}^{XY})$ isolates the impact of the beyond-Born corrections on the deflection field coming from the convergence and $\kappa\kappa\kappa$ bispectrum. Red diamonds mark this contribution in Fig.~\ref{fig:delta-cl-cmb-beyond-born}.
\item $\Delta^{\rm eff, eff-grad}_{\rm eff-1st}(C_{\ell}^{XY})$ isolates the impact of the beyond-Born curl deflection corrections as well as the mixed $\kappa\kappa\omega$ bispectrum. They are shown as blue squares in Fig.~\ref{fig:delta-cl-cmb-beyond-born}.
\item $\Delta^{\rm eff+rotation, eff}_{\rm eff}(C_{\ell}^{XY})$ shows the impact of the polarization rotation only. Fig.~\ref{fig:delta-cl-cmb-beyond-born} shows this correction as pink dots. Here the superscript eff+rotation means that polarization rotation effects have been applied on maps lensed with the $\mathbf{d}^{\rm eff}$ field.
\item $\Delta^{\rm eff+rotation, eff-1st}_{\rm eff-1st}(C_{\ell}^{XY})$ gives the impact of all the beyond-Born corrections with respect to the first-order calculation. These are shown as black dots in Fig.~\ref{fig:delta-cl-cmb-beyond-born}.
\end{enumerate}
We note that the $\Delta^{x, y}_{z}(C_{\ell}^{XY})$ quantities do not account for the effects of the convergence bispectrum due to LSS non-linearities. Because all the deflection fields are extracted from the same fully non-linear matter distribution of the DEMNUni N-Body simulation, they all contain the same amount of non-linearities. They differ only by how we model the bending of the light-ray. Thus, the majority of this contribution of  LSS non-Gaussianity is canceled when computing any $\Delta^{x, y}_{z}(C_{\ell}^{XY})$ quantity.\\*
In Fig.~\ref{fig:delta-cl-cmb-beyond-born} we show the average of the fractional differences of the lensed CMB power spectra for each component of the beyond-Born corrections computed over 100 different realizations of unlensed CMB. The beyond-Born deflection corrections on the power spectra reach the $0.2\%$ level at $\ell\approx 4000$ for TT and TE power spectra. Measuring in detail those corrections is challenging due to their oscillatory shape and our impossibility to run MC simulations on a large set of N-body realizations. However, this is not the case for the corrections on the B-modes power spectrum, because the amplitude of the corrections is larger and reaches the $0.6\%$ level at $\ell\approx 4000$. The amplitude of the beyond-Born convergence correction is consistent in amplitude and shape with the one computed in \citep{lewis-pratten2016, marozzi2016dec} for the angular scales considered in these papers. \\*
The contribution induced by curl-like deflection is negligible for the TT, EE and TE power spectra but is important for the B-modes and accounts for $30\%$ of the signal at small angular scales.  Thus, it cannot be neglected for an accurate analysis of the signal at $\ell\gtrsim 1000$. Moreover, we confirm the findings of \citep{lewis-pratten2016} concerning the impact of the $\kappa\kappa\omega$ correction to the B-modes power spectrum. In particular the difference between $\Delta^{\rm eff, eff-grad}_{\rm eff-1st}(C_{\ell}^{BB})$ and the quantity $C_{\ell}^{BB, \rm eff-curl}/C_{\ell}^{BB, \rm eff-1st}$\footnote{We denote here with $C_{\ell}^{BB, \rm eff-curl}$ and $C_{\ell}^{BB, \rm eff-1st} $ the B-modes power spectrum obtained with $\mathbf{d}^{\rm eff-curl}$ and $\mathbf{d}^{\rm eff-1st}$ respectively.}, which accounts for the curl deflection without the contribution of $\kappa\kappa\omega$, is negligible for all practical purposes as discussed in the next section.
Furthermore, the amplitude of the $\beta^{\rm rotation}$ correction on the polarization power spectra looks negligible up to $\ell\approx 6000$, similarly to the recent analytical findings of \cite{lewis2017}. We postpone a more detailed discussion on this subject to Sec.~\ref{sec:polrot}.\\*
In Fig.~\ref{fig:t-e-modes-theory} and Fig.~\ref{fig:zoom-bmodes} we show a direct comparison of the results of our simulations with the analytical predictions  of \cite{marozzi2016dec} (hereafter MFDD). We found a good agreement between those and our results for the beyond-Born deflection corrections including both the curl and gradient components on all TT, EE and, in particular, for BB power spectra. In Fig.~\ref{fig:t-e-modes-theory} we also show a comparison to the predictions adopting the non-perturbative method of \cite{lewis-pratten2016} for the TT and EE power spectra. This approach leads to a suppression of roughly a factor of 2 of the corrections at small angular scales. However, on the TT power spectrum the difference between the two analytical results is practically unobservable on all scales. On the other hand, on the EE power spectrum, we can clearly observe a better agreement with the non-perturbative theoretical predictions of \cite{lewis-pratten2016} at $\ell\gtrsim 2200$. We note that both the theoretical predictions shown in Fig.~\ref{fig:zoom-bmodes} include all the matter perturbations at $z\leq 1100$, while simulations include contributions from structures located at $z\leq 99$. The good agreement of the theoretical and numerical results shows that the contribution of structures located at $z\geq 99$ to the beyond-Born signal is negligible. \\* 
Because beyond-Born corrections become more important at progressively smaller angular scales, via numerical simulations we were able to investigate them in more detail at $\ell\gg 3000$. At these scales, the signal receives contributions from matter perturbations on scales where the fitting formulae for the matter power spectrum and bispectrum used in the analytic calculations are less accurate. We found that the beyond-Born deflection corrections reach the $1\%$ level at $\ell\approx 6000$ for all TT, EE and BB power spectra.\\*
We finally note that the beyond-Born corrections do not modify the TB and EB power spectrum at a sensible level. This is expected because lensing preserves parity, and we verified that this is true even when beyond-Born corrections are included. 
\noindent
\noindent
\noindent

\subsection{Zooming on lensed B-modes}
The B-modes are the most sensitive signal to beyond-Born corrections and, despite some contribution might become negligible at the power spectrum level, we tried to isolate them on the map level. The most interesting one is the contribution of the mixed bispectrum $\kappa\kappa\omega$ (or, equivalently, $\psi\psi\Omega$). As noted in PL16, two specific contractions of the mixed bispectrum can contribute to the large scales B-modes power spectrum. Following PL16 and adopting the flat-sky approximation plus a second order Taylor expansion around the unlensed direction of Eq.~\eqref{eq:birefringence}, we can write the lensed B-modes in Fourier space as the superposition of different contributions 
\begin{eqnarray}
\tilde{B}(\bell)^{d^{\rm eff}} &=& - \int \frac{d^{2}\bellp}{(2\pi)^{2}}E(\bellp)\sin(2\varphi_{\bell\bellp})\left[\bellp\times\bell\Omega^{\rm eff}(\bell-\bellp) +(\bell-\bellp)\cdot\bellp\psi(\bell-\bellp)\right]\nonumber\\ 
&-&\frac{1}{2}\int\frac{d^{2}\bell_{1}}{(2\pi)^{2}}\int\frac{d^{2}\bell_{2}}{(2\pi)^{2}}\sin(2\varphi_{\bell_{1}\bell})E(\bell_{1})[\bell_{1}\cdot\bell_{2}\psi^{\rm eff}(\bell_{2})+\bell_{1}\times\bell_{2}\Omega^{\rm eff}(\bell_{2})]\nonumber\\
&\times&[\bell_{1}\cdot(\bell_{1}+\bell_{2}-\bell)\psi(\bell-\bell_{1}-\bell_{2})+\bell_{1}\times(\bell_{2}-\bell)\Omega^{\rm eff}(\bell-\bell_{1} -\bell_{2})],
\label{eq:taylor-bl}
\end{eqnarray}
\noindent	
where $\varphi_{\bell\bellp}$ is the angle between the two Fourier modes $\bell$ and $\bellp$. The last equation corrects a sign for the first-order $\Omega(\bell)$ term reported in the published version of PL16 that was later corrected in the latest version of the paper submitted to the arXiv\footnote{version 3 or later at \url{https://arxiv.org/abs/1605.05662}}. We identify with $\tilde{B}(\bell)^{d^{\rm eff-grad}}$, $\tilde{B}(\bell)^{d^{\rm eff-curl}}$, $\tilde{B}(\bell)^{d^{\rm eff-1st}}$ the B-modes obtained setting to zero the $\Omega$ or $\psi$ respectively in the deflection field or using the first-order lensing potential $\psi^{\rm 1st}$, as explained in the previous section. With these quantities in mind and assuming $\psi$ and $\Omega^{\rm eff}$ are nearly Gaussian, we can isolate the two contributions of the mixed bispectrum $\kappa\kappa\omega$ to the B-modes power spectrum computing the following power spectra
\begin{eqnarray}
\Delta^{\psi\Omega}(\bell) &\equiv& B(\bell)^{\rm eff} - B(\bell)^{\rm eff-grad} - B(\bell)^{\rm eff-curl}\\
C_{\ell}^{BB, \rm corr}&=&\langle B(\bell)^{*, \rm eff-grad}B(\bellp)^{\rm eff-curl} \rangle \propto \langle \psi\psi\Omega +\psi\Omega\Omega\nonumber\\
&+&\psi\psi\Omega\Omega\rangle \approx \langle \psi\psi\Omega\rangle + \textrm{higher order}\label{eq:bmodes-correlation}\\
C_{\ell}^{BB, \rm mixed}\equiv C_{\ell}^{BB, \rm \psi\psi\Omega}&=&\langle\Delta^{\psi\Omega}(\bell) ^{*}B(\bellp)^{\rm eff-grad} \rangle \propto
\langle \psi\psi\Omega
 + \psi\Omega\psi\psi\rangle.\label{eq:bmodes-mixed}\\
C_{\ell}^{BB, \psi\Omega\Omega}&=& \langle\Delta^{\psi\Omega}(\bell) ^{*}B(\bellp)^{\rm eff-curl} \rangle \propto
\langle \psi\Omega\Omega
 + \psi\Omega\Omega\Omega\rangle.\label{eq:bmodes-mixed2}
\end{eqnarray}
where we dropped the specific $\bell$-dependencies in the higher order terms to simplify the notation. Following the convention adopted in PL16, we refer to $C_{\ell}^{BB, \rm corr}$ as the correlation term, and to $C_{\ell}^{BB, \rm mixed}$ as the mixed term. In addition to these terms, we isolated an additional contribution to the B-modes, $C_{\ell}^{BB, \psi\Omega\Omega}$, which is related to the $\psi\Omega\Omega$ bispectrum. All these terms are shown in Fig.~\ref{fig:higher-order} together with the B-modes signal induced by the curl deflection alone. $C_{\ell}^{BB, \rm corr}$ and $C_{\ell}^{BB, \rm mixed}$ can be compared with the results shown in Fig.~11 of \cite{pratten2016}. The mixed bispectrum contribution exceeds the signal induced by the curl deflections alone, and displays an amplitude and a characteristic sign inversion at around $\ell\approx 1500$ in agreement with the findings of PL16. The $C_{\ell}^{BB, \rm corr}$ term shows a good agreement with PL16 for $150\leq\ell\leq 1500$. At $\ell\leq150$ we get a much higher contribution with respect to the analytical results but we cannot disentangle if the discrepancy is due to numerical errors in the analytical calculations or boxsize effects in our lightcone construction. The sign of $C_{\ell}^{BB, \rm corr}$ at $1500\lesssim\ell\lesssim 2000$ disagrees with analytical prediction but the discrepancy is driven by the higher order terms contributing to our measurements. The contribution coming from $\psi\Omega\Omega$ bispectrum is in fact non negligible and if we properly subtract it, we recover a qualitative agreement with PL16. The situation is different for $C_{\ell}^{BB, \rm mixed}$. In fact, if we consider $\psi$ and $\Omega$ as Gaussian fields and recall that $C_{\ell}^{\psi\Omega}=0$, the higher order term $\langle\psi^{\rm eff}\Omega^{\rm eff}\psi^{\rm eff}\psi^{\rm eff}$ is approximately 0, providing thus a cleaner measurement of the mixed bispectrum contribution. The same reasoning applies to $C_{\ell}^{BB, \psi\Omega\Omega}$. 
Given their amplitude, we do not expect any of these terms to affect significantly the inflationary B-modes estimation. In fact, they partly cancel at large scales, besides their contribution being further suppressed if a delensing procedure reducing the contribution of the $\psi$-related signal is applied. The dominant contribution that can affect the estimation of $r$ after delensing comes from the presence of curl mode in the deflection. This generates a signal as high as the one produced by primordial tensor perturbations with an amplitude $r\lesssim 10^{-6}$ at $\ell\approx 80$, where the primordial B-modes signal is expected to peak. This level of contamination is far from the sensitivity of any future experiment proposed so far, but might affect any attempt to extract the tilt of the primordial tensor perturbations power spectrum $n_T$ because those measurements require an accurate lensing B-modes subtraction at smaller angular scales. 

\subsection{The impact of polarization rotation}\label{sec:polrot}
\begin{figure}[!htbp]
\centering
\includegraphics[width=.5\textwidth]{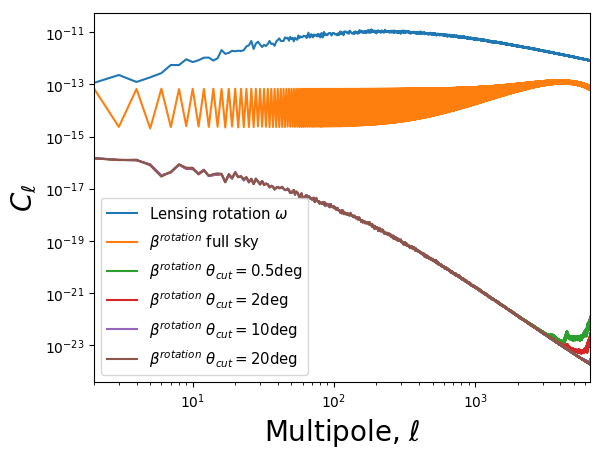}\includegraphics[width=.5\textwidth]{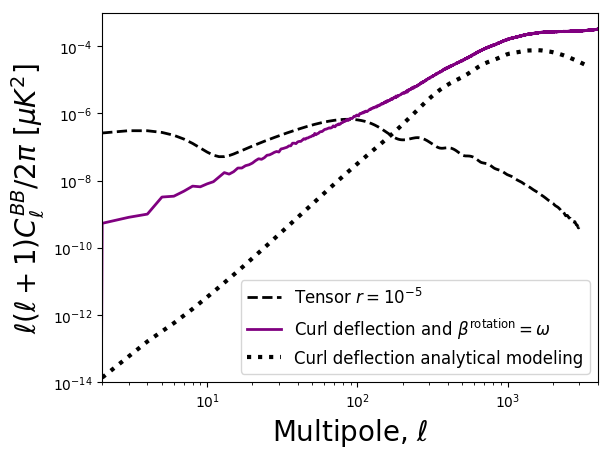}\\
\includegraphics[width=.5\textwidth]{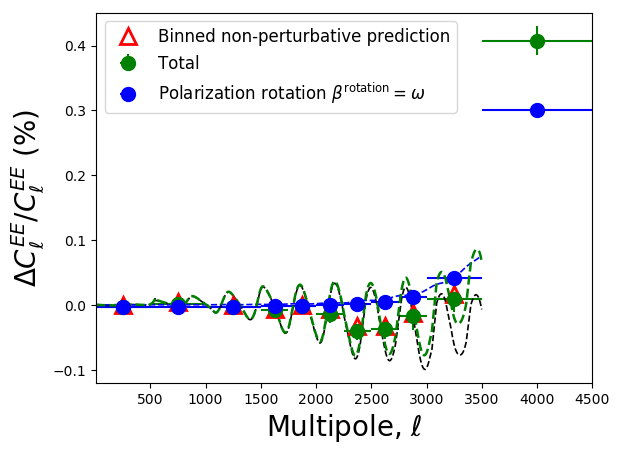}\includegraphics[width=.49\textwidth]{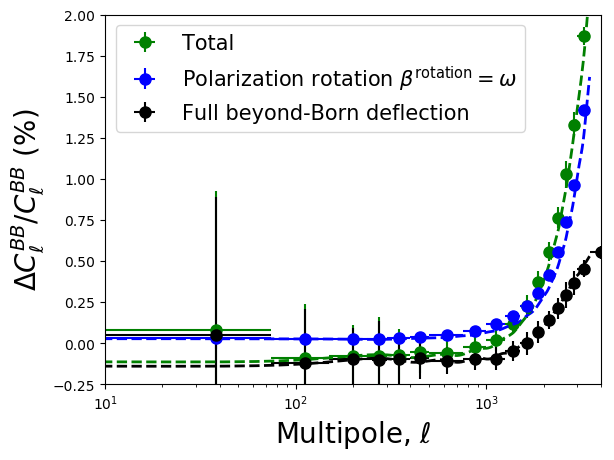}
\caption{Top left: $C_{\ell}^{\beta^{\rm rotation}\beta^{\rm rotation}}$ extracted from the simulation. We show the results computed on the full sky and the one derived masking progressively larger regions around the poles. Because  $C_{\ell}^{\beta^{\rm rotation}\beta^{\rm rotation}} = C_{\ell}^{\omega\omega}$ according to \cite{marozzi2016}, we show $C_{\ell}^{\omega\omega}$ of the simulations for reference. Top right: residual B-modes signal induced by the curl deflection and polarization rotation correction derived from simulations assuming $\beta^{\rm rotation}=\omega$ (purple). In the same panel we show the signal generated by tensor perturbations with $r=10^{-5}$ and by curl deflections alone for reference. 
Bottom: impact of the beyond-Born corrections to the total lensed E-modes (left) and B-modes (right) power spectrum assuming $\beta^{\rm rotation}=\omega$. The results of the simulations are shown as solid dots. The contributions of beyond-Born corrections to the deflection field are shown as a black dashed line, and, in particular, we show the perturbative results for the B-modes (see Fig.~\ref{fig:zoom-bmodes}) and the non-perturbative one for the E-modes (see Fig~\ref{fig:t-e-modes-theory}). The contribution of the polarization rotation alone is shown as a dashed blue line. The total correction corresponding to the sum of  deflection and polarization rotation corrections is shown as a dashed green line. The binned version of this curve is shown with red markers.}
\label{fig:higher-order}
\end{figure}

\begin{figure}[!htb]
\includegraphics[width=.5\textwidth]{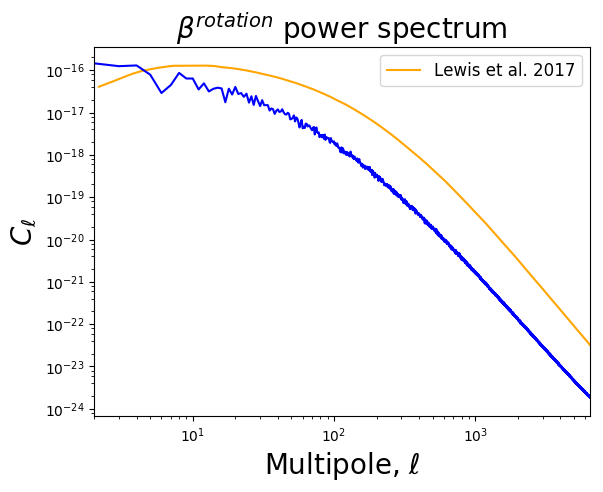}\includegraphics[width=.5\textwidth]{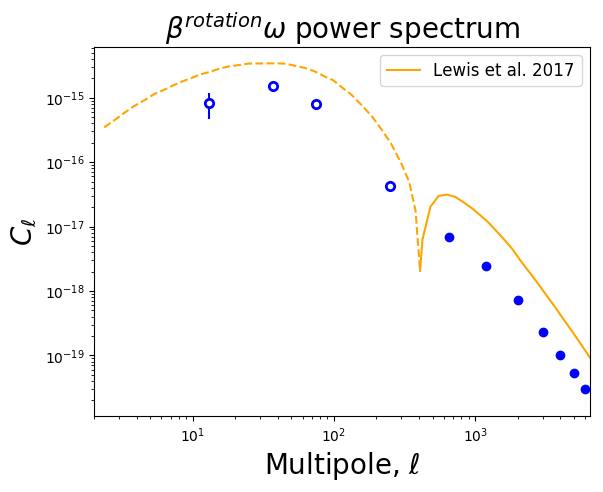}
\caption{Comparison between the predictions of \cite{lewis2017} for $C_{\ell}^{\beta^{\rm rotation}\beta^{\rm rotation}}$ (left) and $C_{\ell}^{\beta^{rotation}\omega}$ cross-correlation (right) with the results of our simulations. Empty dots and dashed line denote negative values. The error bars on the right panel correspond to the error on the mean computed in each bin of multipoles. }
\label{fig:beta-theory-lewis}
\end{figure}

Recently, two independent groups investigated the effect of the polarization rotation corrections on lensed CMB power spectra. MFDD adopted a next-to-leading order formalism in the light-cone gauge, and found that $\beta^{\rm rotation} = \omega$ up to second order in scalar perturbations. This result seems to confirm the guess of \cite{dai2014} that derived the same result if deflections are induced by vector or tensor perturbations at linear order, but speculated on its validity at higher order in scalar perturbations. Conversely, \cite{lewis2017} (hereafter LHC) adopted a different approach based on the geometry of parallel transport, and found that $\beta^{\rm rotation}$ is too small to have a significant impact on lensed CMB power spectra.\\*
In Sec.~\ref{results:power-spectra} we showed that the polarization rotation correction computed using the $\beta^{\rm rotation}$ extracted from our simulations is essentially an unobservable effect.  
In Fig.~\ref{fig:higher-order} we show a comparison of the power spectrum of $\beta^{\rm rotation}$ with the power spectrum of the lensing rotation field $\omega$ extracted from our simulations. These should be equal in the MFDD approach. On the other hand, LHC predicted the power spectrum $\beta^{\rm rotation}$ to have a negative power-law shape and an amplitude between $10^{-16}$ and $10^{-22}$ going from large to small angular scales (see Fig. 6 in \cite{lewis2017} or Fig.~\ref{fig:beta-theory-lewis}). The $C_{\ell}^{\beta^{\rm rotation}\beta^{\rm rotation}}$ of our simulations differs from both these predictions when computed on the full sky. It is in fact dominated by numerical noise at large angular scales but displays an excess of power at $\ell\gtrsim 1000$ approaching the result of MFDD while keeping a systematically lower amplitude. We verified that this result was stable with respect to the choice of the raytracing resolution at 15\% level. However, an accurate inspection of our $\beta^{\rm rotation}$ maps showed us that the technique used to compute $\beta^{\rm rotation}$  produces values of the polarization rotation angle roughly 1000 times higher than the typical r.m.s of the map for pixels located close to the poles. This effect is clearly unphysical, because the $\beta^{\rm rotation}$ field should be isotropic. Such systematic effect at the poles generated ringing in harmonic domain, and was the responsible of the highly oscillatory behaviour of the full sky $C_{\ell}^{\beta^{\rm rotation}\beta^{\rm rotation}}$. In order to mitigate these artifacts, we decided to mask the pixels having colatitude $0<\theta<\theta_{\rm cut}$ and $\theta_{\rm cut}< \theta < \pi -  \theta_{\rm cut}$ when computing $C_{\ell}^{\beta^{\rm rotation}\beta^{\rm rotation}}$. As it can be seen in Fig.~\ref{fig:higher-order}, once we masked a modest sky fraction around the poles (e.g.  $\theta_{\rm cut} = 0.5 { \rm deg}$ ), $C_{\ell}^{\beta^{\rm rotation}\beta^{\rm rotation}}$ changed significantly. This value remained stable at $\ell\lesssim 4000$ if we masked larger areas around the poles. This new result disagrees clearly with the results of MFDD. On the other hand, in Fig.~\ref{fig:beta-theory-lewis} we show a comparison between this new trend of $C_{\ell}^{\beta^{\rm rotation}\beta^{\rm rotation}}$ and the predictions of LHC. Despite a difference in the amplitude of $\beta^{\rm rotation}$ of roughly one order of magnitude, our measurement of $C_{\ell}^{\beta^{\rm rotation}\beta^{\rm rotation}}$  shows a qualitative agreement with the shape of LHC results. We also recover with good agreement the prediction of the cross-correlation power spectrum between $\beta^{\rm rotation}$ and $\omega$ (see Fig.~\ref{fig:beta-theory-lewis}). Despite this good qualitative agreement, the fact that the results are sensitive to the details of the processing of the maps suggests that a more precise measurement of $\beta^{\rm rotation}$ might require a redefinition of the technique used to isolate this effect. We plan to investigate this problem in more detail in future work. \\*
Keeping in mind the caveat just outlined and for the sake of completeness, we also checked explicitly the validity of the calculations of MFDD for the lensed CMB power spectra. To this purpose, we rotated the lensed CMB simulations described in the previous sections by the lensing rotation angle $\omega$ extracted from the simulations, instead of $\beta^{\rm rotation}$. In Fig.~\ref{fig:higher-order} we show that this might indeed become the dominant effect on the power spectrum level, and could increase the amplitude of the correction to 6\% on the E-modes and to 11\% on the B-modes power spectrum at $\ell\approx 6000$. In Fig.~\ref{fig:higher-order} we also show a comparison of these results with the theoretical prediction derived in MFDD for the polarization rotation correction. In this scenario, the residual contribution to the large scale B-modes increased with respect to the results shown in Fig.~\ref{fig:zoom-bmodes}, and could reach the level of $r\approx 10^{-5}$. This value is still below any foreseeable experimental target.
We stress that this test does not validate the $\beta^{\rm rotation}=\omega$ result of MFDD (which is disfavoured in light of the results discussed above), but only their calculations of the lensed CMB power spectra at the correct perturbative level if $\beta^{\rm rotation}=\omega$.\\*
Finally we note that the polarization rotation correction does not contribute to any isotropic parity breaking that would create non-zero TB and EB correlations, because both $\omega$ and $\beta^{\rm rotation}$ have effectively zero mean\footnote{An analogous mechanism can arise in the context of CMB polarization experiments if the orientation of polarized detectors on the sky is misestimated \cite{keating2013}.}.

\subsection{Detection perspectives}\label{sec:detection}
The future generation of CMB experiments, such as CMB-S4, will produce high resolution maps of CMB polarization over a large fraction of the sky (nearly 50\%). Assuming the level of signal extracted from our simulations, these corrections might not be negligible \cite{marozzi2016dec-param}. We expect that for an experiment similar to \cite{cmbs4}, with $1.4\mu\textrm{K-arcmin}$ polarization sensitivity, 1 arcmin FWHM angular resolution, and observing a range of angular scales $30\leq \ell\leq 5000$, beyond-Born correction on TT, EE and BB power spectra could be measured with a cumulative signal to noise ratio (S/N) of 4.8, 0.5, 0.7 respectively, if we consider both the $\beta^{\rm rotation}$ and deflection corrections extracted from the simulations (as well as the same input cosmology used for the N-body simulation\footnote{We assumed a Gaussian covariance for the B-modes power spectrum degraded by a scale-independent factor 1.2 to take into account the non-Gaussian component of the covariance introduced by weak lensing \cite{benoit-levy2012}).}. Even in the most optimistic case where $\beta^{\rm rotation}=\omega$ (and thus the impact of the total beyond-Born corrections on the polarization power spectra is higher), the S/N for the corrections on EE and BB power spectra would be 1.4 and 2.2, respectively. Thus, in practice, a robust detection of post-Born corrections on the lensed CMB power spectra will be challenging as will highly depend on the capability to control the impact of extragalactic foreground to the required level of precision and accuracy. This is true especially for the TT power spectrum, where both the signal and the extragalactic contaminations are more important. We note that a realistic forecast of the detectability of these corrections is also complicated by the uncertainties in the modeling of the non-linear matter power spectrum and by degeneracies with other cosmological parameters affecting the CMB small-scale power (see e.g. \cite{marozzi2016dec-param} for a discussion on post-Born correction and $N_{eff}$ measurement).\\*
A direct detection of the corrections on the $C_{\ell}^{\kappa\kappa}$, or a direct reconstruction of the lensing curl potential $\Omega$, will be extremely challenging for upcoming experiments \cite{pratten2016, sheere2016}. For the curl potential extracted from the simulations, in particular, the cumulative signal to noise for its direct reconstruction will only be 1.2, assuming a simplified Gaussian reconstruction noise as in \cite{namikawa2012}. 
Given the amplitude of the effect measured in the simulation, the detection of any $\beta^{\rm rotation}$ contribution on lensed CMB power spectra is not feasible. We discuss this issue in more detail in appendix \ref{appendix:beta}.

%% file: conclusions.tex

\section{Conclusions}\label{sec:concl}
In this paper we presented an improvement of the raytracing algorithm through N-body simulations for CMB lensing presented in \cite{Calabrese14}. The modified method propagates the full CMB lensing jacobian, deflecting light rays trajectories using a similar multiple-lens approach. The improved numerical efficiency of the algorithm made possible a detailed analysis of the second-order lensing effects due to the relaxation of the Born approximation on the convergence, shear and lensing rotation fields at arcsecond resolution on the full sky. We also improved the setup of the N-body simulation of \cite{Calabrese14} employing a simulation with larger boxsize and resolution included in the DEMNUni suite \citep{castorina2015,carbone2016}. \\*
We tested the robustness of our method against resolution effects in the raytracing and in the N-body simulation used in this work, as well as against the number of employed lens planes finding no evidence of numerical artifacts. \\*
We showed that the effect of beyond-Born corrections on the statistical properties of both the $\kappa$ and $\omega$ field 1-point PDF and power spectra, as well as on non-Gaussian statistics. We found a good agreement with theoretical predictions of \cite{pratten2016} for the beyond-Born $C_{\ell}^{\kappa\kappa}$ and $C_{\ell}^{\omega\omega}$ power spectra and a qualitative confirmation that the beyond-Born corrections tend to suppress the amount of non-Gaussianity in the $\kappa$ field. However, we decided to postpone a quantitative characterization of the $\kappa\kappa\kappa$ and $\kappa\kappa\omega$ bispectrum to future works. \\*
We then used the lensing observables extracted from our raytracing simulation to evaluate the effect of beyond-Born corrections on the lensed CMB power spectra up to $\ell\approx 6000$. We compared these findings with recent analytical predictions including beyond-Born deflection corrections of \cite{marozzi2016dec, lewis-pratten2016}, finding a very good agreement in the range of angular scales considered in that work ($\ell\lesssim 3500$) in particular for the B-modes power spectrum. We also found that the non-perturbative  
approach of \cite{lewis-pratten2016}, which predicts a suppression of the amplitude of beyond-Born corrections on lensed temperature and E-modes power spectra at $\ell\gtrsim 2000$, describes the results of the simulation better than the calculations of \cite{marozzi2016dec}. 
We showed, for the first time in the literature, the impact of the $\kappa\omega\omega$ bispectrum on the lensed B-modes power spectrum, and found it to be comparable to the contribution of the curl-mode in the deflection field at $\ell\gtrsim 1000$.  \\*
In addition, we measured the second-order gravitational rotation of CMB polarization, $\beta^{\rm rotation}$ directly from the simulations. Our results indicate that the impact of $\beta^{\rm rotation}$ on lensed CMB power spectra is negligible as found recently by \cite{lewis2017}.  
We compared $C_{\ell}^{\beta^{\rm rotation}\beta^{\rm rotation}}$ and the cross-correlation $C_{\ell}^{\beta^{\rm rotation}\omega}$ measured in our simulations with theoretical predictions of \cite{marozzi2016dec} and \cite{lewis2017}. 
Despite some differences in amplitude for both these observables still remain, we found a good qualitative agreement with the prediction of \cite{lewis2017}. We recovered the shape and sign inversion of $C_{\ell}^{\beta^{\rm rotation}\omega}$ and found a good qualitative agreement with the shape and amplitude of $C_{\ell}^{\beta^{\rm rotation}\beta^{\rm rotation}}$ once we masked the region around the poles of the $\beta^{\rm rotation}$ map in the power spectrum computation. These pixels present in fact an anomalously high contribution which we interpreted as a consequence of a limitation of our approach to extract $\beta^{\rm rotation}$. Because this effect is small and a direct reconstruction of $\beta^{\rm rotation}$ field is essentially unfeasible, we decided to postpone the design of a more accurate simulation method of $\beta^{\rm rotation}$ to future work. \\* 
Finally, we discussed the possibility to detect the beyond-Born signatures with future high sensitivity and high resolution CMB polarization experiments, such as CMB-S4. We showed that the corrections to the temperature power spectrum could be detected with good significance, although the extragalactic foregrounds will act as the main (probably definitive) obstacle. The situation for the detection of beyond-Born corrections to E and B-modes power spectra is conversely more challenging as the expected signal to noise is less than 2.